\documentclass[aps,prapplied,twocolumn,floatfix,amsmath,amssymb,longbibliography]{revtex4-2}
\usepackage{graphicx}
\usepackage{xcolor}
\usepackage[normalem]{ulem}
\usepackage{gensymb}
\usepackage{textcomp}
\usepackage{bm}
\usepackage{placeins}
\usepackage{float}

\begin{document}
\title{Composite Dielectric-Loaded Metallic Metagratings for\\Dual-Polarized Transmissive Beam Manipulation}
\author{Vitali Specktor}
\email{vitali.sp@campus.technion.ac.il}
\author{Yuval Singer}
\author{Ilya Varenisov}
\author{Ariel Epstein}
\email{epsteina@ee.technion.ac.il}
\affiliation{Andrew and Erna Viterbi Faculty of Electrical and Computer Engineering, Technion--Israel Institute of Technology, Haifa 3200003, Israel}

\begin{abstract}
We present and experimentally demonstrate a comprehensive analytical suite for the design of transmissive metal-based metagratings (MGs) for dual-polarized anomalous refraction. Considering a composite configuration of perforated metallic slabs with controlled dielectric loading, we develop a rigorous mode-matching formulation tying the various MG degrees of freedom to the various reflected and transmitted Floquet-Bloch modes, enabling efficient beam deflection for both transverse electric (TE) and transverse magnetic (TM) polarizations. Inherently versatile, the analytical framework accurately models single-layer and double-layer constructs with multiple rectangular perforations per unit cell, each potentially loaded with dielectric material. We harness this degree-of-freedom abundance to devise diverse polarization-sensitive transmissive MGs, ranging from all-metallic single-layer and double-layer structures for single-polarized (TM or TE) perfect anomalous refraction, to dielectric-loaded architectures implementing dual-polarized routing of the incident wave. Beyond highlighting concrete paths to address TE-polarized fields in such formations, our results, validated via full-wave simulations and experiments at K band, showcase the great potential of metal-based MGs for manipulating fields in transmission, facilitating effective synthesis of durable dual-polarized beamforming superstrates for environmentally demanding and high power antenna applications.
\end{abstract}

\maketitle

\section{Introduction}
\label{sec:introduction}
The quest for precise electromagnetic wave manipulation has driven significant advances in complex media science and technology, relying on intricately assembled subwavelength polarizable particles (meta-atoms) to achieve extreme and exotic field-moulding functionalities.
Among these developments, metagratings (MGs) have emerged as a particularly promising class of devices, featuring low-profile, sparse, typically periodic meta-atom arrangements \cite{Radi2022}.
Despite the reduced complexity relative to the more common densely-populated metasurface solutions, this alternative platform has been used in recent years to successfully demonstrate highly efficient and diverse beamforming capabilities. Unlike conventional gratings, commonly utilizing periodic modulations of material properties, MGs exploit the resonant behavior and coupling characteristics of carefully designed meta-atoms to obtain the reported unprecedented control over diffraction patterns and wave steering \cite{Radi2017, Sell2017, Wong2018, paniagua2018metalens}. The key to achieve this is the utilization of reliable analytical models, relating, based on  Floquet-Bloch (FB) theory, the MG scattering coefficients directly to their underlying structural geometric parameters. This enables accurate consideration of near- and far-field interactions between individual meta-atoms, orchestrating their response as to eventually yield through interference the target scattered field distribution (i.e., prescribed FB coupling coefficients).

At microwave frequencies, the majority of reported MG implementations have utilized printed circuit board (PCB) technologies, where metallic strips or patches on dielectric substrates serve as the fundamental scattering elements \cite{Memarian2017, Epstein2017_1, Rabinovich2018, Popov2018, Wong2018, Rabinovich2020, Popov2020_1, Xu2021_1}. In particular for loaded-strip-based configurations, adaptations of wire-media analytical models \cite{Wait1954, Tretyakov2003, Epstein2017_1} provided a rigorous framework for rapid design of versatile fabrication-ready PCB MGs, facilitating realization of a wide range of MGs for beam deflection \cite{Rabinovich2018,Popov2018,Rabinovich2020,Popov2020_1,Popov2021, Xu2021_1, Yi2023}, focusing \cite{Rabinovich2020,Xu2021_1}, waveguide mode manipulation \cite{Killamsetty2021, Biniashvili2022, Tan2025}, polarization control \cite{Shklarsh2024, Elad2024, Nguyen2024}, absorption \cite{Yashno2022, Boust2022, Tan2023}, and antenna enhancement \cite{Popov2020, Xu2022, Xu2022_2, Kerzhner2022}, to name a few. 

While this approach yields highly efficient devices and offers considerable design flexibility as well as straightforward manufacturing, the reliance on thin dielectric substrates introduces inherent limitations. In communication or radar systems employed in demanding scenarios, such dielectrics may be continuously exposed to thermal gradients, mechanical stress, and environmental conditions, which may challenge their structural robustness.
In addition, the thin conductive traces and narrow interlayer spacings typically limit the ability to handle high electromagnetic field intensities, rendering PCB-based MGs less suitable for high-power applications \cite{Zhang2019, Ahmed2022, Pourgholamhossein2025}. Furthermore, dielectric losses tend to grow with frequency, which may reduce the effectiveness of such PCB devices for beamforming at the millimeter-wave and sub-terahertz regimes \cite{quevedo2018, gonzalez2018, quevedo2025}. In contrast, metallic-based configurations usually feature enhanced mechanical durability, enable improved power handling, and often offer inherently lower losses and greater fabrication precision (e.g., through advanced machining or additive manufacturing techniques) relevant for high-frequency operation. Correspondingly, considering these benefits of metallic configurations -- many times making them the preferable choice for applications such as space-mounted antennas, high-power radar installations, and high-frequency communication systems \cite{Chahat2018,Zhang2019,Ahmed2022-full,Quevedo2022,Quevedo2023} -- attempts to devise metallic-based MGs have been carried out in recent years, striving to adapt the appealing synthesis approach to relevant formations while retaining the trademark versatility and beamforming efficiency of this class of complex media.

Indeed, recent years have seen numerous proposals for metallic-based MGs \cite{Rabinovich2020_1, Rahmanzadeh2020, Hemmatyar2021, Rajabalipanah2021, Rajabalipanah2022, Rahmanzadeh2022}, considering metallic slabs with subwavelength trenches, grooves, or through-holes (with or without dielectric loading) acting as the basic scattering unit (meta-atom). However, these implementations are currently restricted to reflective configurations or single-polarization operation, and do not systematically address dual-polarized transmissive beam manipulation -- an essential capability if MGs are to be used as effective beamforming superstrates. More recently, in a theoretical work, Alex-Amor \textit{et al.} \cite{AlexAmor2023} have explored penetrable 3D metallic configurations with connected air cavities of different dimensions, cascaded judiciously to tune the spectral content of a dual-polarized incident signal. Nonetheless, the formulation considered specific rectangular and hard waveguide element arrangement in the cascade, with overall subwavelength lateral periodicity, also including simplifying assumptions neglecting high-order evanescent modes at certain interfaces. Consequently, such structures can only manipulate the fundamental FB harmonics (i.e., affect only specular reflection or direct transmission), suitable mainly for polarization control and frequency-filtering applications. To allow wavefront manipulation in transmission through an accurately designed MG construct, a scheme allowing analysis and engineering of a more general composite featuring larger periodicities and increased number of lateral degrees of freedom -- while enabling efficient reflection suppression and effective interaction with both polarizations simultaneously -- is required.

In this work, we address these gaps, developing a comprehensive analytical framework for the synthesis of dual-polarized transmissive metallic MGs, based on rigorous mode-matching techniques and FB theory \cite{Specktor2025}. Augmenting the formalism developed in \cite{Rabinovich2020_1} for reflective all-metallic MGs, we extend the analysis to through-perforated metallic slabs with multiple rectangular apertures per period, providing analytical access to the coupling coefficients of both transverse electric (TE) and transverse magnetic (TM), reflected and transmitted, FB modes associated with a given MG configuration. Through the direct relations between the structural parameters -- number of perforations, their arrangement within the period, and the dimensions of each of these waveguide sections -- and the scattering coefficients, an inverse problem can be readily formulated, yielding the optimal MG layout for prescribed polarized beam deflection requirements.

To overcome the inherent challenges of TE-polarized beam manipulation in such planar metallic devices \cite{Rabinovich2020_1}, we further expand the formulation to include additional degrees of freedom. First, we allow dielectric loading of the perforations, increasing the number of propagating modes within the subwavelength waveguides -- previously identified as a core parameter for enabling advanced diffraction engineering in waveguide-based MGs \cite{Dong2020}. Second, we derive the scattering solution for cascaded metallic MG pairs, facilitating additional tuning of the FB mode coupling coefficients through multimodal multiple reflections \cite{Marcus2022}.

As demonstrated by diverse designs validated in full-wave simulations and selectively via experimental K-band prototypes, the developed methodology enables through these extensions the realization of effective dual-polarized transmissive metallic-based MGs. In particular, we synthesize and verify numerically single-layer and double-layer all-metallic MGs for single-polarized (TM and TE, respectively) anomalous refraction; as well as two different single-layer dielectric-loaded composites for dual-polarized operation, exhibiting polarization-independent or polarization-sensitive beam steering. The dielectric-loaded designs -- relatively demanding fabrication-wise -- are further manufactured and experimentally characterized at K band, confirmed to deliver high 
diffraction efficiencies in laboratory measurements. Overall, these results establish a foundation for systematic semianalytical development of high-performance dual-polarized MG devices suitable for transmissive applications, enabling integration of these versatile beamforming superstrates in applications requiring polarization diversity and robust operation, such as spaceborne antennas, high-power radars, and high-frequency communication systems.

\section{Theory}
\label{sec:theory}
\subsection{Single-Layer Configuration}
\label{subsec:single_layer}
\begin{figure*}[!t]
\centerline{\includegraphics[width=2\columnwidth]{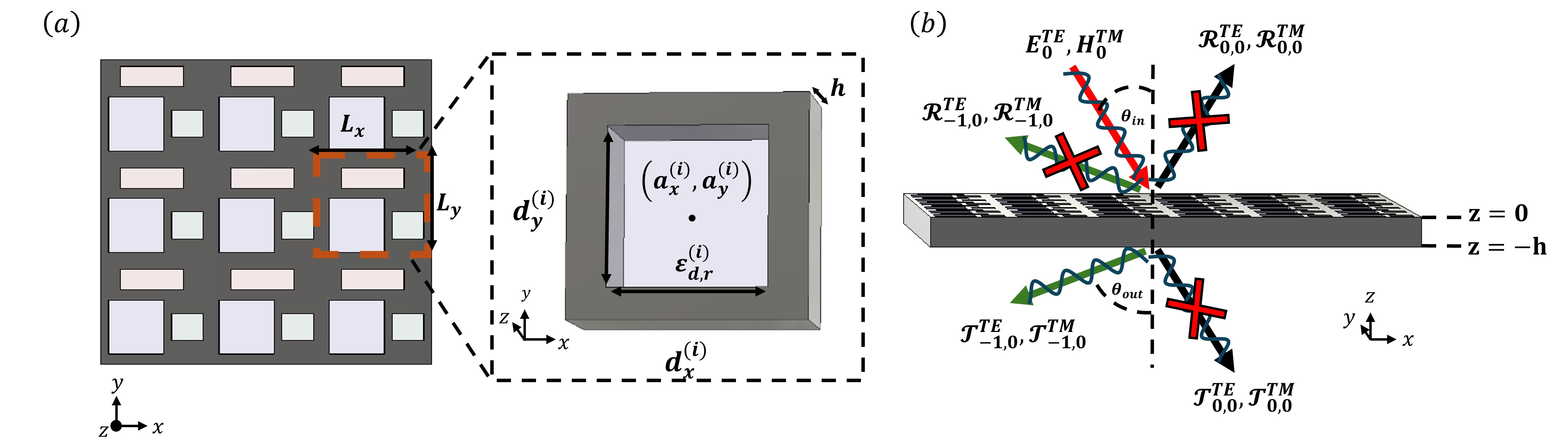}}
\caption{Physical configuration of the considered transmissive metallic-based MG. (a) Top view; the MG consists of a metallic slab (gray) of thickness $h$ featuring multiple rectangular perforations, arranged periodically according to the highlighted $L_x\times L_y$ unit cell (dashed orange rectangle). Inset: detailed close-up of the cross-section of the $i$th perforation, with center coordinates $(a_x^{(i)}, a_y^{(i)})$ and lateral dimensions $d_x^{(i)}\times d_y^{(i)}$, potentially filled with dielectric material of permittivity $\varepsilon_{d,r}^{(i)}$. (b) Side view of a single-layer MG occupying the region $-h<z<0$, intended to funnel the incident power carried by a dual-polarized plane-wave excitation ($E_0^{\text{TE}}$, $H_0^{\text{TM}}$) from $\theta_{\text{in}}$, into the anomalous refraction order $(-1,0)$ (transmitted towards $\theta_{\text{out}}$), while suppressing all other scattered propagating modes; $\mathcal{R}_{\mathbf{n}}$ and $\mathcal{T}_{\mathbf{n}}$ indicate, respectively, the reflection and transmission modal amplitudes of the ${\mathbf{n}}=(n_x,n_y)$-order FB harmonics, for either TE- and TM-polarized scattered fields.}
\label{fig:geometry}
\end{figure*} 
We consider a waveguide-based MG configuration, consisting of periodically arranged rectangular through-perforations (potentially filled with dielectric material) within a metallic slab of thickness $h$ made of perfect electric conductor (PEC), as illustrated in Fig. \ref{fig:geometry}. The periodicities along $x$ and $y$ are, respectively, $L_x$ and $L_y$, with each $L_x\times L_y$ unit cell containing $N_p$ rectangular perforations. The $i$th perforation is characterized by its \emph{relative} dielectric permittivity $\varepsilon_{d,r}^{(i)}$, lateral dimensions $d_x^{(i)} \times d_y^{(i)}$, and center coordinates $(a_x^{(i)}, a_y^{(i)})$ [see inset of Fig. \ref{fig:geometry}(a)], defining its boundaries as $x_\pm^{(i)} = a_x^{(i)} \pm d_x^{(i)}/2$ and $y_\pm^{(i)} = a_y^{(i)} \pm d_y^{(i)}/2$. A dual-polarized plane wave of angular frequency $\omega$ impinges on the structure from the upper half-space ($z > 0$), at an angle of incidence $\theta_{\text{in}}$ within the $\widehat{xz}$ plane [Fig. \ref{fig:geometry}(b)]. The $y$ components of the incident electric and magnetic fields are given by
\begin{align}
\begin{aligned}
\!\!\!E_{y,\text{inc}}^{\text{TE}} = E_0^{\text{TE}} e^{-jk_x x} e^{jk_z z};\,
H_{y,\text{inc}}^{\text{TM}} = H_0^{\text{TM}} e^{-jk_x x} e^{jk_z z},
\end{aligned}
\label{equ:incident_fields}
\end{align}
where $k_x = k\sin\theta_{\text{in}}$ and $k_z = k\cos\theta_{\text{in}}$ denote, respectively, the transverse and longitudinal components of the wave vector, with $k = \omega\sqrt{\mu_0\varepsilon_0}$ being the free-space wavenumber. $E_0^{\text{TE}}$ and $H_0^{\text{TM}}$ are the complex amplitudes of the TE ($E_z=0$) and TM ($H_z=0$) polarization components of the incident wave, analyzed separately according to the superposition principle. 

As discussed in Section \ref{sec:introduction}, our goal is to establish an effective methodology for the synthesis of metallic-based dual-polarized beam-deflecting MGs, exercising meticulous control over the coupling of the incoming wave to each of the diffraction orders (FB modes). In particular, as a fundamental beamforming functionality, we would be interested in funneling the incident power from the fundamental $(0,0)$-order towards the prescribed anomalous refraction harmonic $(-1,0)$, addressing either TE-polarized or TM-polarized waves, or both. To this end, we first stipulate the fields in all regions of the corresponding scattering problem [Fig. \ref{fig:geometry}(b)]. Within each of the $N_p$ dielectric-loaded perforations (acting as truncated rectangular waveguides), the electromagnetic fields are expanded in terms of the respective waveguide eigenmodes. Specifically, for the $i$th perforation ($-h<z<0$), the $z$-components of the electric and magnetic fields read
\begin{align}
\begin{aligned}
\!\!E_{z,\mathrm{prf}}^{\text{TM},(i)} \!\!&=\!\! \sum_{m_x=1}^{\infty} \sum_{m_y=1}^{\infty} \!\!\eta_d^{(i)} \!\!\left( \mathcal{A}_{+,\mathbf{m}}^{\text{TM},(i)} e^{-jk_{z,d}^{(i)} z} + \mathcal{A}_{-,\mathbf{m}}^{\text{TM},(i)} e^{jk_{z,d}^{(i)} z} \right) \\
& \times \sin\left[\frac{\pi m_x}{d_x^{(i)}}(x - x_-^{(i)})\right] \sin\left[\frac{\pi m_y}{d_y^{(i)}}(y - y_-^{(i)})\right]; \\
\!\!H_{z,\mathrm{prf}}^{\text{TE},(i)} \!\!&=\!\! \sum_{m_x^*=0}^{\infty} \sum_{m_y^*=0}^{\infty} \left( \mathcal{A}_{+,\mathbf{m}^*}^{\text{TE},(i)} e^{-jk_{z,d}^{(i)} z} + \mathcal{A}_{-,\mathbf{m}^*}^{\text{TE},(i)} e^{jk_{z,d}^{(i)} z} \right) \\
& \times \cos\left[\frac{\pi m_x^*}{d_x^{(i)}}(x - x_-^{(i)})\right] \cos\left[\frac{\pi m_y^*}{d_y^{(i)}}(y - y_-^{(i)})\right],
\end{aligned}
\label{equ:perforation_fields}
\end{align}
where we denoted the waveguide mode indices as $\mathbf{m} \equiv (m_x, m_y)$ and $\mathbf{m}^* \equiv (m_x^*, m_y^*)$, for brevity; the asterisk notation $\mathbf{m}^*$ indicates that the mode indices $m_x^*$ and $m_y^*$ must not simultaneously vanish for TE modes \cite{Pozar2012}. The coefficients $\mathcal{A}_{\pm,\mathbf{m}}^{\text{TM},(i)}$ and $\mathcal{A}_{\pm,\mathbf{m}^*}^{\text{TE},(i)}$ represent the complex amplitudes of forward ($+$) and backward ($-$) propagating modes, with the modal propagation constant within the dielectric-loaded perforation given by $[k_{z,d}^{(i)}]^2 = k^2 \varepsilon_{d,r}^{(i)} - [\pi m_x/d_x^{(i)}]^2 - [\pi m_y/d_y^{(i)}]^2$; the corresponding wave impedance is $\eta_d^{(i)} = [\mu_0/\varepsilon_0\varepsilon_{d,r}^{(i)}]^{1/2}$. 

In the regions external to the metallic structure ($z > 0$ and $z < -h$), the scattered fields are formulated using FB expansions, consistent with diffraction theory for periodic configurations \cite{Bhattacharyya2006}. Correspondingly, the reflected fields ($z > 0$) are expressed as
\begin{align}
\begin{aligned}
\!\!\!E_{z,\text{ref}}^{\text{TM}} &= \!\!\!\!\!\sum_{n_x=-\infty}^{\infty} \sum_{n_y=-\infty}^{\infty} \eta\mathcal{R}_{\mathbf{n}}^{\text{TM}}  
e^{-jk_x^{(n_x)} x} e^{-jk_y^{(n_y)} y} e^{-jk_z z}; \\
\!\!\!H_{z,\text{ref}}^{\text{TE}} &= \!\!\!\!\!\sum_{n_x=-\infty}^{\infty} \sum_{n_y=-\infty}^{\infty} \mathcal{R}_{\mathbf{n}}^{\text{TE}}  
e^{ -jk_x^{(n_x)} x} e^{-jk_y^{(n_y)} y} e^{-jk_z z},
\end{aligned}
\label{equ:reflected_fields}
\end{align}
while the transmitted fields ($z < -h$) are given by
\begin{align}
\begin{aligned}
\!\!\!E_{z,\text{tra}}^{\text{TM}} &= \!\!\!\!\sum_{n_x=-\infty}^{\infty} \sum_{n_y=-\infty}^{\infty} \eta\mathcal{T}_{\mathbf{n}}^{\text{TM}} 
e^{-jk_x^{(n_x)} x} e^{-jk_y^{(n_y)} y} e^{jk_z z}; \\
\!\!\!H_{z,\text{tra}}^{\text{TE}} &= \!\!\!\!\sum_{n_x=-\infty}^{\infty} \sum_{n_y=-\infty}^{\infty} \mathcal{T}_{\mathbf{n}}^{\text{TE}} 
e^{-jk_x^{(n_x)} x} e^{-jk_y^{(n_y)} y} e^{jk_z z},
\end{aligned}
\label{equ:transmitted_fields}
\end{align}
where, again, we use a compact form to denote the FB mode indices $\mathbf{n} \equiv (n_x, n_y)$; $\eta = (\mu_0/\varepsilon_0)^{1/2}$ is the free-space wave impedance. The complex variables $\mathcal{T}_{\mathbf{n}}^{\text{TM}}$ ($\mathcal{T}_{\mathbf{n}}^{\text{TE}}$) and $\mathcal{R}_{\mathbf{n}}^{\text{TM}}$ ($\mathcal{R}_{\mathbf{n}}^{\text{TE}}$) stand for the TM (TE) polarized modal amplitudes of transmitted or reflected waves, respectively, associated with the $(n_x,n_y)$-order FB mode. The transverse wavenumbers are $k_x^{(n_x)} = k\sin\theta_{\text{in}} + (2\pi/L_x)n_x$ and $k_y^{(n_y)} = (2\pi/L_y)n_y$, while the longitudinal wavenumber follows from the dispersion relation, $k_z^2 = k^2 - [k_x^{(n_x)}]^2 - [k_y^{(n_y)}]^2$. Substituting the longitudinal field components in Eqs. \eqref{equ:perforation_fields}-\eqref{equ:transmitted_fields} into Maxwell's equations allows derivation of their tangential counterparts \cite{Balanis2012}, enabling formulation of the complete electromagnetic field vectors at each point in our configuration.

Once the expressions for the fields everywhere have been formally established, we proceed as a second step to derive analytical relations between the MG's geometric parameters [Fig. \ref{fig:geometry}(a)] and its scattering characteristics [Fig. \ref{fig:geometry}(b)], facilitating the solution of the inverse problem (MG synthesis, \textit{cf.} Section \ref{subsec:synthesis}). This is achieved by enforcing the continuity of the tangential electric and magnetic field components at the two MG/air interfaces ($z=0$ and $z=-h$), employing mode matching \cite{Rabinovich2020_1} to establish rigorous relationships between the unknown modal amplitudes within the perforations ($\mathcal{A}_{\pm,\mathbf{m}}^{\text{TM},(i)}$ and $\mathcal{A}_{\pm,\mathbf{m}^*}^{\text{TE},(i)}$) and the scattering coefficients in the external regions ($\mathcal{R}_{\mathbf{n}}^{\text{TM}}$, $\mathcal{R}_{\mathbf{n}}^{\text{TE}}$, $\mathcal{T}_{\mathbf{n}}^{\text{TM}}$, and $\mathcal{T}_{\mathbf{n}}^{\text{TE}}$). 

As in \cite{Rabinovich2020_1}, to allow practical calculations, the infinite sums in Eqs. \eqref{equ:perforation_fields}-\eqref{equ:transmitted_fields} must first be truncated to a finite number of terms. Specifically, we limit the waveguide mode orders to $M_x$ and $M_y$, while the FB diffraction orders towards $x$ and $y$ are retained up to a maximum of $|n_x|\leq N_x/2$ and $|n_y|\leq N_y/2$, respectively ($N_x$ and $N_y$ being even integers). Utilizing yet another compact notation for the truncation limits $\mathbf{M} \equiv (M_x, M_y)$ and $\mathbf{N} \equiv (N_x, N_y)$, summations over waveguide mode indices [Eq. \eqref{equ:perforation_fields}] range from $\mathbf{m} = \mathbf{1}$ to $\mathbf{M}$ for TM modes, or $\mathbf{m}^* = \mathbf{0}$ to $\mathbf{M}$ for TE modes; while summations over FB harmonic indices [Eqs. \eqref{equ:reflected_fields}-\eqref{equ:transmitted_fields}] span from $\mathbf{n} = -\mathbf{N}/2$ to $\mathbf{N}/2$. Thus, considering $(N_x+1)\times(N_y+1)$ FB modal coefficients for each of the two polarizations, in both reflection ($\mathcal{R}_{\mathbf{n}}^{\text{TM}}$ and $\mathcal{R}_{\mathbf{n}}^{\text{TE}}$) and transmission ($\mathcal{T}_{\mathbf{n}}^{\text{TM}}$ and $\mathcal{T}_{\mathbf{n}}^{\text{TE}}$); as well as backward and forward $M_x\times M_y$ TM- ($\mathcal{A}_{\pm,\mathbf{m}}^{\text{TM},(i)}$) and $M_x\times M_y+M_x+M_y$ TE- ($\mathcal{A}_{\pm,\mathbf{m}^*}^{\text{TE},(i)}$) waveguide eigenmode amplitudes for each of the $N_p$ perforations; this truncation scheme yields a total of $U=4(N_x+1)(N_y+1)+2N_p(2M_xM_y+M_x+M_y)$ unknown coefficients we need to resolve (for a given MG configuration) through an appropriate set of $U$ linear equations.

To establish this desired system of equations, we extend the procedure in \cite{Rabinovich2020_1} to the transmissive case considered herein. We begin by applying the continuity conditions for the tangential \emph{electric} field at both the top and bottom interfaces of the perforated slab, and use inner products with the suitable eigenmodes to relate the various modal coefficients to one another. Specifically, we multiply these electric-field boundary conditions [derived from Eqs. \eqref{equ:incident_fields}-\eqref{equ:transmitted_fields}] by the FB harmonics $e^{jk_x^{(n_x)} x} e^{jk_y^{(n_y)} y}$ and integrate over their orthogonality domain, i.e., the unit cell area $L_x\times L_y$ \cite{Rabinovich2020_1}. For a given FB diffraction order $\mathbf{n}=(n_x,n_y)$, performing this for the equations associated with the continuity of the $x$-component of the electric field at the top ($z=0$) and bottom ($z=-h$) interfaces, respectively, yields (Appendix \ref{appendix:coefficients})
\begin{equation}
\begin{split}
&\!\!\!\left(\alpha^{\text{TM}(\text{I})}_{\mathbf{n}}\mathcal{R}_{\mathbf{n}}^{\text{TM}}+\alpha^{\text{TE}(\text{I})}_{\mathbf{n}}\mathcal{R}_{\mathbf{n}}^{\text{TE}}\right)\\
&\!\!\!-\sum_{i}^{N_p}\frac{\psi^{(i)}_{n,m}}{L_x L_y}
\Bigg[
\sum_{\mathbf{m}^*=0}^{\mathbf{M}}\Big(\beta^{\text{TE}(\text{I},i)}_{\mathbf{m}^*}\mathcal{A}_{+,\mathbf{m}^*}^{\text{TE},(i)}+ \beta^{\text{TE}(\text{I},i)}_{\mathbf{m}^*}\mathcal{A}_{-,\mathbf{m}^*}^{\text{TE},(i)} \Big)\\
&\!\!\!\quad+\sum_{\mathbf{m}=1}^{\mathbf{M}}\Big(\beta^{\text{TM}(\text{I},i)}_{\mathbf{m}}\mathcal{A}_{+,\mathbf{m}}^{\text{TM},(i)}- \beta^{\text{TM}(\text{I},i)}_{\mathbf{m}}\mathcal{A}_{-,\mathbf{m}}^{\text{TM},(i)} \Big)
\Bigg]=-S^{(\text{I})}_\mathbf{n}
\end{split}
\label{eq:Ex_upper}
\end{equation}
and
\begin{equation}
\begin{split}
&\!\!\!\left(-\alpha^{\text{TM}(\text{I})}_{\mathbf{n}}\mathcal{T}_{\mathbf{n}}^{\text{TM}} +\alpha^{\text{TE}(\text{I})}_{\mathbf{n}}\mathcal{T}_{\mathbf{n}}^{\text{TE}}\right)e^{-jk_z h}\\
&\!\!\!-\sum_{i}^{N_p}\frac{\psi^{(i)}_{\mathbf{n},\mathbf{m}}}{L_x L_y}
\Bigg[\sum_{\mathbf{m}^*=\mathbf{0}}^{\mathbf{M}}\Big(\beta^{\text{TE}(\text{I},i)}_{\mathbf{m}^*}
\mathcal{A}_{+,\mathbf{m}^*}^{\text{TE},(i)}e^{jk_{z,d}^{(i)}h}\\
&\!\!\!\quad\quad\quad\quad\quad\quad\quad\quad\quad+ \beta^{\text{TE}(\text{I},i)}_{\mathbf{m}^*}\mathcal{A}_{-,\mathbf{m}^*}^{\text{TE},(i)}e^{-jk_{z,d}^{(i)}h} \Big)\\
&\!\!\!\quad\quad\quad+\sum_{\mathbf{m}=\mathbf{1}}^{\mathbf{M}}\Big(\beta^{\text{TM}(\text{I},i)}_{\mathbf{m}}\mathcal{A}_{+,\mathbf{m}}^{\text{TM},(i)}e^{jk_{z,d}^{(i)}h}\\
&\!\!\!\quad\quad\quad\quad\quad\quad- \beta^{\text{TM}(\text{I},i)}_{\mathbf{m}}\mathcal{A}_{-,\mathbf{m}}^{\text{TM},(i)}e^{-jk_{z,d}^{(i)}h} \Big)
\Bigg]=0.
\end{split}
\label{eq:Ex_lower}
\end{equation}
Similarly, the relations derived for the same $\mathbf{n}$ from the continuity conditions corresponding to the $y$-component of the electric field read
\begin{equation}
\begin{split}
&\left(\alpha^{\text{TM}(\text{II})}_{\mathbf{n}}\mathcal{R}_{\mathbf{n}}^{\text{TM}}+\alpha^{\text{TE}(\text{II})}_{\mathbf{n}}\mathcal{R}_{\mathbf{n}}^{\text{TE}}\right)\\
&-\sum_{i}^{N_p}\frac{\chi^{(i)}_{\mathbf{n},\mathbf{m}}}{L_x L_y}
\Bigg[
\sum_{\mathbf{m}^*=0}^{\mathbf{M}}\Big(\beta^{\text{TE}(\text{II},i)}_{\mathbf{m}^*}\mathcal{A}_{+,\mathbf{m}^*}^{\text{TE},(i)}
+ \beta^{\text{TE}(\text{II},i)}_{\mathbf{m}^*}\mathcal{A}_{-,\mathbf{m}^*}^{\text{TE},(i)} \Big)\\
&\quad+\sum_{\mathbf{m}=1}^{\mathbf{M}}\Big(\beta^{\text{TM}(\text{II},i)}_{\mathbf{m}}\mathcal{A}_{+,\mathbf{m}}^{\text{TM},(i)}- \beta^{\text{TM}(\text{II},i)}_{\mathbf{m}}\mathcal{A}_{-,\mathbf{m}}^{\text{TM},(i)} \Big)
\Bigg]=-S^{(\text{II})}_\mathbf{n}
\end{split}
\label{eq:Ey_upper}
\end{equation}
for the top interface ($z=0$), and
\begin{equation}
\begin{split}
&\left(-\alpha^{\text{TM}(\text{II})}_{\mathbf{n}}\mathcal{T}_{\mathbf{n}}^{\text{TM}} +\alpha^{\text{TE}(\text{II})}_{\mathbf{n}}\mathcal{T}_{\mathbf{n}}^{\text{TE}}\right)e^{-jk_z h}\\
&-\sum_{i}^{N_p}\frac{\chi^{(i)}_{\mathbf{n},\mathbf{m}}}{L_x L_y}
\Bigg[
\sum_{\mathbf{m}^*=0}^{\mathbf{M}}\Big(\beta^{\text{TE}(\text{II},i)}_{\mathbf{m}^*}\mathcal{A}_{+,\mathbf{m}^*}^{\text{TE},(i)}e^{jk_{z,d}^{(i)}h}\\
&\quad\quad\quad\quad\quad\quad\quad\quad\quad+ \beta^{\text{TE}(\text{II},i)}_{\mathbf{m}^*}\mathcal{A}_{-,\mathbf{m}^*}^{\text{TE},(i)}e^{-jk_{z,d}^{(i)}h} \Big)\\
&\quad\quad\quad+\sum_{\mathbf{m}=1}^{\mathbf{M}}\Big(\beta^{\text{TM}(\text{II},i)}_{\mathbf{m}}\mathcal{A}_{+,\mathbf{m}}^{\text{TM},(i)}e^{jk_{z,d}^{(i)}h}\\
&\quad\quad\quad\quad\quad\quad- \beta^{\text{TM}(\text{II},i)}_{\mathbf{m}}\mathcal{A}_{-,\mathbf{m}}^{\text{TM},(i)}e^{-jk_{z,d}^{(i)}h} \Big)
\Bigg]=0
\end{split}
\label{eq:Ey_lower}
\end{equation}
for the interface at $z=-h$.

As expected, coupling between the $\mathbf{n}$th FB and $\mathbf{m}$th waveguide modes is governed by the corresponding overlap integrals over the $i$th perforation, $\psi^{(i)}_{\mathbf{n},\mathbf{m}}$ and $\chi^{(i)}_{\mathbf{n},\mathbf{m}}$, featuring different expressions for $E_x$-continuity [Eqs. \eqref{eq:Ex_upper}-\eqref{eq:Ex_lower}] and $E_y$-continuity [Eqs. \eqref{eq:Ey_upper}-\eqref{eq:Ey_lower}] mode matching due to different lateral field profiles (see Appendix \ref{appendix:coefficients}). These differences also affect the functional form of the $\alpha_{\mathbf{n}}^{\text{TE}(q)}$, $\alpha_{\mathbf{n}}^{\text{TM}(q)}$ and $\beta_{\mathbf{m}^*}^{\text{TE}(q,i)}$, $\beta_{\mathbf{m}}^{\text{TM}(q,i)}$ coefficients (provided in Appendix \ref{appendix:coefficients} as well), emerging systematically when the transverse components of the electromagnetic fields are derived from the longitudinal ones, which explains the need for different superscripts $(\text{I})$ and $(\text{II})$ in Eqs. \eqref{eq:Ex_upper}-\eqref{eq:Ex_lower} and \eqref{eq:Ey_upper}-\eqref{eq:Ey_lower}, respectively. The source terms $S_{\mathbf{n}}^{(\text{I})}$ and $S_{\mathbf{n}}^{(\text{II})}$ represent the contribution of the incident fundamental $\mathbf{0}=(0,0)$-order FB mode to the right-hand side of the respective boundary conditions, vanishing for $\mathbf{n}\neq\mathbf{0}$ (\textit{cf.} Table \ref{tab:coefficients} and Eq. \eqref{eq:source_terms} in Appendix \ref{appendix:coefficients}). Overall, repeating the projection procedure yielding Eqs. \eqref{eq:Ex_upper}-\eqref{eq:Ey_lower} for the entire set of FB modes results in $4(N_x+1)(N_y+1)$ linear equations, associated with the two tangential electric field components at the two relevant interfaces.

To complement these as to reach the required set of $U$ equations [corresponding to the $U$ unknowns in Eqs. \eqref{equ:perforation_fields}-\eqref{equ:transmitted_fields}], we proceed to enforce the continuity conditions for the tangential \emph{magnetic} fields. As pointed out in \cite{Rabinovich2020_1}, resolving these requires special care due to the magnetic field discontinuity at the metallic surfaces covering portions of the MG facets (see Fig. \ref{fig:geometry}), arising from surface currents induced on the PEC. Consequently, continuity of these magnetic field components can only be enforced over the perforation apertures. Thus, the mode-matching procedure in this case necessitates projection of the magnetic field boundary conditions onto the waveguide modal basis functions and integration over each perforation aperture \emph{separately}; here, as opposed to Eqs. \eqref{eq:Ex_upper}-\eqref{eq:Ey_lower}, FB mode orthogonality will not be of use, since the inner product cannot be performed on the entire unit-cell domain (FB mode orthogonality domain). 

Correspondingly, we first multiply the continuity condition associated with the $x$-component of the magnetic field by $\sin[(\pi m_x/d_x^{(i)})(x - x_-^{(i)})]\cos[(\pi m_y/d_y^{(i)})(y - y_-^{(i)})]$ (where $m_x \neq 0$ ensures that the condition does not trivially vanish) and integrate over the cross-section of the $i$th waveguide aperture. In analogy to Eqs. \eqref{eq:Ex_upper}-\eqref{eq:Ey_lower}, for a given perforation index $i$ and waveguide mode order $\mathbf{m}=(m_x,m_y)$, two relations are formed when performing this procedure at the $z=0$ and $z=-h$ planes, yielding, respectively,
\begin{equation}
\begin{split}
&\sum_{\mathbf{n}=-\mathbf{N}/2}^{\mathbf{N}/2}
  \left(\alpha^{\text{TM}(\text{III})}_{\mathbf{n}}\mathcal{R}_{\mathbf{n}}^{\text{TM}}
  +\alpha^{\text{TE}(\text{III})}_{\mathbf{n}}\mathcal{R}_{\mathbf{n}}^{\text{TE}}\right)
  \chi^{*(i)}_{\mathbf{n},\mathbf{m}}\\
&-\frac{d_x^{(i)} d_y^{(i)}}{4}(1+\delta_{m_y,0})
  \Big(\beta^{\text{TE}(\text{III},i)}_{\mathbf{m}^*}\mathcal{A}_{+,\mathbf{m}^*}^{\text{TE},(i)}
  -\beta^{\text{TE}(\text{III},i)}_{\mathbf{m}^*}\mathcal{A}_{-,\mathbf{m}^*}^{\text{TE},(i)}\\
&\quad\quad+\beta^{\text{TM}(\text{III},i)}_{\mathbf{m}}\mathcal{A}_{+,\mathbf{m}}^{\text{TM},(i)}
  +\beta^{\text{TM}(\text{III},i)}_{\mathbf{m}}\mathcal{A}_{-,\mathbf{m}}^{\text{TM},(i)}\Big)
  =-S_{\mathbf{m}} ^{(\text{III},i)}
\end{split}
\label{eq:Hx_upper}
\end{equation}
and
\begin{equation}
\begin{split}
&\sum_{\mathbf{n}=-\mathbf{N}/2}^{\mathbf{N}/2}
  \left(\alpha^{\text{TM}(\text{III})}_{\mathbf{n}}\mathcal{T}_{\mathbf{n}}^{\text{TM}}
  -\alpha^{\text{TE}(\text{III})}_{\mathbf{n}}\mathcal{T}_{\mathbf{n}}^{\text{TE}}\right)
  e^{-jk_z h}\chi^{*(i)}_{\mathbf{n},\mathbf{m}}\\
&-\frac{d_x^{(i)} d_y^{(i)}}{4}(1+\delta_{m_y,0})
  \Big(\beta^{\text{TE}(\text{III},i)}_{\mathbf{m}^*}\mathcal{A}_{+,\mathbf{m}^*}^{\text{TE},(i)}e^{jk_{z,d}^{(i)}h}\\
&\quad\quad-\beta^{\text{TE}(\text{III},i)}_{\mathbf{m}^*}\mathcal{A}_{-,\mathbf{m}^*}^{\text{TE},(i)}e^{-jk_{z,d}^{(i)}h}
  +\beta^{\text{TM}(\text{III},i)}_{\mathbf{m}}\mathcal{A}_{+,\mathbf{m}}^{\text{TM},(i)}e^{jk_{z,d}^{(i)}h}\\
&\quad\quad+\beta^{\text{TM}(\text{III},i)}_{\mathbf{m}}\mathcal{A}_{-,\mathbf{m}}^{\text{TM},(i)}e^{-jk_{z,d}^{(i)}h}\Big)
  =0
\end{split}
\label{eq:Hx_lower}
\end{equation}
with the $\alpha_{\mathbf{n}}^{\text{TE}(q)}$, $\alpha_{\mathbf{n}}^{\text{TM}(q)}$ and $\beta_{\mathbf{m}^*}^{\text{TE}(q,i)}$, $\beta_{\mathbf{m}}^{\text{TM}(q,i)}$ coefficients relevant to these boundary conditions ($q=\text{III}$) defined, similar to Eqs. \eqref{eq:Ex_upper}-\eqref{eq:Ey_lower}, in Appendix \ref{appendix:coefficients}. Since $m_x\neq 0$ while $m_y$ ranges from $0$ to $M_y$, each of the sets in Eqs. \eqref{eq:Hx_upper} and \eqref{eq:Hx_lower} corresponds to $M_x\times (M_y+1)$ independent relations per perforation, amounting to $2N_p\times(M_x\times M_y+M_y)$ equations in total. 

Next, we apply this technique to the conditions demanding the continuity of the $y$-component of the magnetic field. Considering the different eigenmode basis (different boundary conditions within the waveguide) relevant for this component, we execute the inner product with the corresponding $\cos[(\pi m_x/d_x^{(i)})(x - x_-^{(i)})]\sin[(\pi m_y/d_y^{(i)})(y - y_-^{(i)})]$ (with $m_y \neq 0$ to prevent trivial satisfaction) and integrate (again) over the perforation apertures \cite{Rabinovich2020_1}. For given perforation index $i$ and mode order $\mathbf{m}=(m_x,m_y)$, this leads to
\begin{equation}
\begin{split}
&\sum_{\mathbf{n}=-\mathbf{N}/2}^{\mathbf{N}/2}
  \left(\alpha^{\text{TM}(\text{IV})}_{\mathbf{n}}\mathcal{R}_{\mathbf{n}}^{\text{TM}}
  +\alpha^{\text{TE}(\text{IV})}_{\mathbf{n}}\mathcal{R}_{\mathbf{n}}^{\text{TE}}\right)
  \psi^{*(i)}_{\mathbf{n},\mathbf{m}}\\
&-\frac{d_x^{(i)} d_y^{(i)}}{4}(1+\delta_{m_x,0})
  \Big(\beta^{\text{TE}(\text{IV},i)}_{\mathbf{m}^*}\mathcal{A}_{+,\mathbf{m}^*}^{\text{TE},(i)}
  -\beta^{\text{TE}(\text{IV},i)}_{\mathbf{m}^*}\mathcal{A}_{-,\mathbf{m}^*}^{\text{TE},(i)}\\
&\quad\quad+\beta^{\text{TM}(\text{IV},i)}_{\mathbf{m}}\mathcal{A}_{+,\mathbf{m}}^{\text{TM},(i)}
  +\beta^{\text{TM}(\text{IV},i)}_{\mathbf{m}}\mathcal{A}_{-,\mathbf{m}}^{\text{TM},(i)}\Big)
  =-S_{\mathbf{m}}^{(\text{IV},i)}
\end{split}
\label{eq:Hy_upper}
\end{equation}
and
\begin{equation}
\begin{split}
&\sum_{\mathbf{n}=-\mathbf{N}/2}^{\mathbf{N}/2}
  \left(\alpha^{\text{TM}(\text{IV})}_{\mathbf{n}}\mathcal{T}_{\mathbf{n}}^{\text{TM}}
  -\alpha^{\text{TE}(\text{IV})}_{\mathbf{n}}\mathcal{T}_{\mathbf{n}}^{\text{TE}}\right)
  e^{-jk_z h}\psi^{*(i)}_{\mathbf{n},\mathbf{m}}\\
&-\frac{d_x^{(i)} d_y^{(i)}}{4}(1+\delta_{m_x,0})
  \Big(\beta^{\text{TE}(\text{IV},i)}_{\mathbf{m}^*}\mathcal{A}_{+,\mathbf{m}^*}^{\text{TE},(i)}e^{jk_{z,d}^{(i)}h}\\
&\quad\quad-\beta^{\text{TE}(\text{IV},i)}_{\mathbf{m}^*}\mathcal{A}_{-,\mathbf{m}^*}^{\text{TE},(i)}e^{-jk_{z,d}^{(i)}h}
  +\beta^{\text{TM}(\text{IV},i)}_{\mathbf{m}}\mathcal{A}_{+,\mathbf{m}}^{\text{TM},(i)}e^{jk_{z,d}^{(i)}h}\\
&\quad\quad+\beta^{\text{TM}(\text{IV},i)}_{\mathbf{m}}\mathcal{A}_{-,\mathbf{m}}^{\text{TM},(i)}e^{-jk_{z,d}^{(i)}h}\Big)
  =0
\end{split}
\label{eq:Hy_lower}
\end{equation}
for the top ($z=0$) and bottom ($z=-h$) interfaces, respectively. 

Reviewing the derivation of these magnetic-field-related boundary conditions, we note that the factor $(1+\delta_{m_y,0})$ [$(1+\delta_{m_x,0})$] in Eqs. \eqref{eq:Hx_upper}-\eqref{eq:Hx_lower} [\eqref{eq:Hy_upper}-\eqref{eq:Hy_lower}] arises from the integral of the cosine eigenfunction over the perforation aperture \cite{Balanis2012}. Furthermore, $S_{\mathbf{m}}^{(\text{III},i)}$ and $S_{\mathbf{m}}^{(\text{IV},i)}$  [defined in Appendix \ref{appendix:coefficients} alongside their electric-field-related counterparts from Eqs. \eqref{eq:Ex_upper}-\eqref{eq:Ey_lower}] stem from the excitation of the $\mathbf{m}$th waveguide mode in the $i$th perforation due to the incident fundamental $(0,0)$-order FB mode; accordingly, their magnitude is governed by the respective overlap integrals $\chi^{(i)}_{\mathbf{0},\mathbf{m}}$ and $\psi^{(i)}_{\mathbf{0},\mathbf{m}}$. Similar to Eqs. \eqref{eq:Hx_upper}-\eqref{eq:Hx_lower}, since $m_y\neq 0$ while $m_x$ ranges from $0$ to $M_x$, Eqs. \eqref{eq:Hy_upper}-\eqref{eq:Hy_lower} together contribute $2N_p\times(M_x\times M_y+M_y)$ equations. 

Overall, combining the $4(N_x+1)(N_y+1)$ equations formed by Eqs. \eqref{eq:Ex_upper}-\eqref{eq:Ey_lower} with the $2N_p(2M_x M_y+M_x+M_y)$ equations formed by Eqs. \eqref{eq:Hx_upper}-\eqref{eq:Hy_lower} yields a set of $U$ independent linear equations with $U$ unknowns $\mathcal{R}_{\mathbf{n}}^{\text{TM/TE}}$, $\mathcal{T}_{\mathbf{n}}^{\text{TM/TE}}$, $\mathcal{A}_{\pm,\mathbf{m}}^{\text{TM},(i)}$ and $\mathcal{A}_{\pm,\mathbf{m}^*}^{\text{TE},(i)}$ [introduced in Eqs. \eqref{equ:perforation_fields}-\eqref{equ:transmitted_fields}], which can be cast into a matrix form:
\begin{equation}
\mathbf{Z}_{U\times U} \mathbf{I}_{U\times 1} = \mathbf{V}_{U\times 1}.
\label{eq:matrix_system}
\end{equation}
In Eq. \eqref{eq:matrix_system}, $\mathbf{Z}_{U\times U}$ is the impedance matrix, whose elements are constructed from the modal coefficients, namely $\alpha_{\mathbf{n}}^{\text{TE}(q)}$, $\alpha_{\mathbf{n}}^{\text{TM}(q)}$ for the $\mathbf{n}$th FB mode and $\beta_{\mathbf{m}^*}^{\text{TE}(q,i)}$, $\beta_{\mathbf{m}}^{\text{TM}(q,i)}$ for the $\mathbf{m}$th waveguide mode in the $i$th perforation, and the overlap integrals $\psi^{(i)}_{\mathbf{n},\mathbf{m}}$ and $\chi^{(i)}_{\mathbf{n},\mathbf{m}}$; 
$\mathbf{I}_{U\times 1}$ is the vector of unknown field amplitudes (comprising the reflected and transmitted field components $\mathcal{R}_{\mathbf{n}}^{\text{TM/TE}}$, $\mathcal{T}_{\mathbf{n}}^{\text{TM/TE}}$, and the waveguide modal amplitudes $\mathcal{A}_{\pm,\mathbf{m}}^{\text{TM},(i)}$, $\mathcal{A}_{\pm,\mathbf{m}^*}^{\text{TE},(i)}$); and $\mathbf{V}_{U\times 1}$ is the excitation vector containing the incident wave amplitudes and source terms $S_{\textbf{m}}^{(q)}$ and $S_{\textbf{n}}^{(q)}$; explicit block-matrix expressions for $\mathbf{Z}$, $\mathbf{I}$, and $\mathbf{V}$ appear in Appendix \ref{appendix:coefficients} as well. Therefore, for a given MG geometry [Fig. \ref{fig:geometry}(a)], the fields scattered due to the incident wave [Fig. \ref{fig:geometry}(b)] may be obtained through standard matrix inversion, $\mathbf{I} = \mathbf{Z}^{-1}\mathbf{V}$. From a synthesis perspective -- as shall be discussed in Section \ref{subsec:synthesis} and demonstrated in Section \ref{sec:results} -- this result and its underlying model Eqs. \eqref{equ:incident_fields}-\eqref{eq:Hy_lower} may be subsequently leveraged for resolving the inverse problem: identifying those MG design specifications that would lead to desired (prescribed) scattering characteristics.

\subsection{Double-Layer Configuration}
\label{subsec:double_layer}
\begin{figure}[!t]
\centerline{\includegraphics[width=1.0\columnwidth]{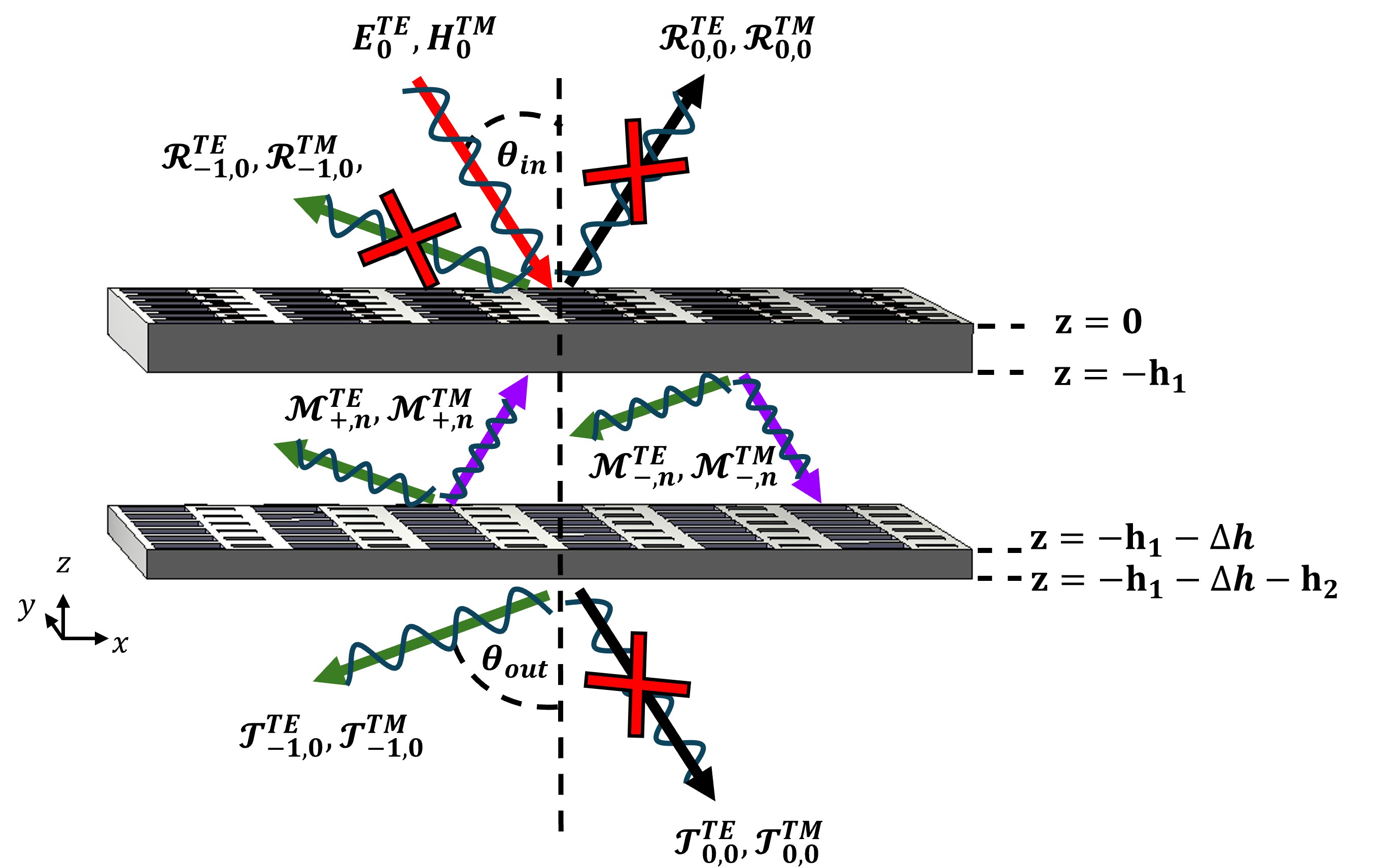}}
\caption{Side view of a double-layer MG (Section \ref{subsec:double_layer}), consisting of two perforated metallic layers of thicknesses $h_1$ and $h_2$ [each with unit cells as described in Fig. \ref{fig:geometry}(a)], separated by an air gap of thickness $\Delta h$ (total thickness $h = h_1 + \Delta h + h_2$). The composite MG is intended to funnel the incident power carried by a dual-polarized plane wave excitation ($E_0^{\text{TE}}$, $H_0^{\text{TM}}$) from $\theta_{\text{in}}$, into the anomalous refraction order $(-1,0)$ (transmitted towards $\theta_{\text{out}}$), while suppressing all other scattered propagating modes. 
In addition to the reflection and transmission modal amplitudes $\mathcal{R}_{\mathbf{n}}$ and $\mathcal{T}_{\mathbf{n}}$ of the various FB harmonics 
[see also Fig. \ref{fig:geometry}(b) and Section \ref{subsec:single_layer}], the intermediate air gap supports forward and backward propagating FB harmonics $\mathcal{M}_{\pm,\mathbf{n}}$ coupling the two metallic layers.}
\label{fig:double_layer}
\end{figure}

To enhance the design flexibility and provide additional degrees of freedom for achieving specific (challenging) scattering characteristics (further discussed in Section \ref{subsec:te_double}), we extend the analytical framework of Section \ref{subsec:single_layer} to accommodate double-layer MG configurations, as depicted in Fig. \ref{fig:double_layer}. This structure consists of two perforated (PEC) metallic slabs of the type described in Fig. \ref{fig:geometry}(a), separated by an air gap of height $\Delta h$. The top and bottom slabs share the same unit cell periodicities $L_x$ (along $x$) and $L_y$ (along $y$), while their thicknesses and number of perforations per period may differ, denoted as $h_1$ and $h_2$, and $N_{p,1}$ and $N_{p,2}$, respectively. The introduction of the intermediate air region enables additional control over the FB scattering phenomena, affecting the coupling coefficients to and from the truncated waveguide meta-atoms through multimodal multiple reflections \cite{Marcus2022}, thereby expanding the accessible design space for dual-polarized anomalous refraction \cite{Dong2020}. 

Overall, the scattering analysis for this augmented scenario follows the rationale and outline presented in Section \ref{subsec:single_layer}; in the following, we highlight the main differences between the single- and double-layer formulations, indicating the modifications needed to obtain the scattered fields relevant to Fig. \ref{fig:double_layer}. To begin with, the fields within the perforations of each layer are described using the same waveguide modal expansions introduced in Eq. \eqref{equ:perforation_fields}, but now distinguished by an additional layer index $l$. Specifically, the modal amplitudes are denoted as $\mathcal{A}_{\pm,\mathbf{m}^*}^{\text{TE},(i),(l)}$ and $\mathcal{A}_{\pm,\mathbf{m}}^{\text{TM},(i),(l)}$, representing the forward and backward propagating TE and TM modes, respectively, in the $i$th perforation of the $l$th layer, where $l=1$ ($l=2$) designates the top (bottom) metallic layer. 

Next, we move to address the fields in the intermediate air region ($-h_1 \geq z \geq -h_1-\Delta h$), which were absent in the single-layer formulation. Since in this region, the fields are being scattered due to interactions with two $L_x\times L_y$-periodic configurations (from above and below); and since the excitation from above consists of corresponding FB harmonics [see Fig. \ref{fig:geometry}(b) and Eq. \eqref{equ:transmitted_fields} of Section \ref{subsec:single_layer}]; the fields in the air gap -- which must also satisfy the constraints imposed by the FB theorem -- share the same FB modal basis as the reflected and transmitted waves. Consequently, these fields are expressed as superpositions of propagating and evanescent FB harmonics traveling in both the forward ($+z$) and backward ($-z$) directions:
\begin{equation}
\begin{split}
E_{z,\text{mid}}^{\text{TM}} &=\!\!\!\! \sum_{n_x=-\infty}^{\infty}
  \sum_{n_y=-\infty}^{\infty} \!\!\!\!\eta \left[ \mathcal{M}_{+,\mathbf{n}}^{\text{TM}}e^{-jk_z z}\right.\\
&\quad\quad\quad\quad\left.+\mathcal{M}_{-,\mathbf{n}}^{\text{TM}}e^{jk_z z} \right]
  e^{-jk_x^{(n_x)} x} e^{-jk_y^{(n_y)} y} ; \\
H_{z,\text{mid}}^{\text{TE}} &=\!\!\!\! \sum_{n_x=-\infty}^{\infty}
  \sum_{n_y=-\infty}^{\infty} \!\!\!\!\left[ \mathcal{M}_{+,\mathbf{n}}^{\text{TE}}e^{-jk_z z}\right.\\
&\quad\quad\quad\left.+\mathcal{M}_{-,\mathbf{n}}^{\text{TE}}e^{jk_z z} \right]
  e^{-jk_x^{(n_x)} x} e^{-jk_y^{(n_y)} y} ,
\end{split}
\label{equ:middle_fb}
\end{equation}
where $\mathcal{M}_{+,\mathbf{n}}^{\text{TM}}$ and $\mathcal{M}_{-,\mathbf{n}}^{\text{TM}}$ ($\mathcal{M}_{+,\mathbf{n}}^{\text{TE}}$ and $\mathcal{M}_{-,\mathbf{n}}^{\text{TE}}$) denote the complex amplitudes of the downward-propagating and upward-propagating TM (TE) polarized $(n_x,n_y)$-order FB harmonics in the intermediate region, respectively; the transverse and longitudinal wavenumbers remain as defined in the single-layer formulation [\textit{cf.} Eqs. \eqref{equ:reflected_fields}-\eqref{equ:transmitted_fields}]. 

Together with the incident ($z>0$), reflected ($z>0$), and transmitted ($z<-h$) fields, defined exactly as before [Eqs. \eqref{equ:incident_fields},  \eqref{equ:reflected_fields}, and \eqref{equ:transmitted_fields}, respectively], the extension of Eq. \eqref{equ:perforation_fields} and the introduced Eq. \eqref{equ:middle_fb} enable stipulation of the fields in the entire double-layer MG domain (Fig. \ref{fig:double_layer}), with the tangential field components obtained from these via substitution into Maxwell's equations, as noted in Section \ref{subsec:single_layer} \cite{Balanis2012}. Importantly, the entire field solution is uniquely characterized by the associated set of field amplitudes, with the intermediate region FB modes of Eq. \eqref{equ:middle_fb} corresponding to $\mathcal{M}_{\pm,\mathbf{n}}^{\text{TM/TE}}$ contributing additional $4(N_x+1)(N_y+1)$ unknowns compared to the single-layer problem; and the number of waveguide modal amplitudes increased now to $2(N_{p,1}+N_{p,2})(2M_xM_y+M_x+M_y)$, in consistency with the total number of perforations in both layers and the number of $\mathcal{A}_{\pm,\mathbf{m}}^{\text{TM},(i),(l)}$ and $\mathcal{A}_{\pm,\mathbf{m}^*}^{\text{TE},(i),(l)}$ instances in the double-layer version of Eq. \eqref{equ:perforation_fields}. Overall, to resolve the fields in the double-layer configuration, it is required to determine the values of the resulting $U_{\mathrm{dbl}}=8(N_x+1)(N_y+1) + 2(N_{p,1}+N_{p,2})(2M_xM_y+M_x+M_y)$ unknowns.

Similar to the single-layer case in Section \ref{subsec:single_layer}, this resolution is obtained by employing an analogous mode-matching procedure, with continuity conditions enforced at each interface by projecting onto the appropriate modal basis functions. However, the double-layer geometry introduces two additional interfaces compared to the single-layer structure, resulting in a larger system of equations. Specifically, boundary conditions must now be satisfied at four distinct planes: $z = 0$ (top face of the first layer), $z = -h_1$ (bottom of the first layer/top boundary of the air gap), $z = -h_1-\Delta h$ (bottom boundary of the air gap/top face of the second layer), and $z = -h_1-\Delta h-h_2 = -h$ (bottom of the second layer). At the two intermediate interfaces ($z = -h_1$ and $z = -h_1-\Delta h$), the fields in the waveguide perforations must be matched to the FB expansion in the air gap, coupling the modal amplitudes in the two metallic layers through the intermediate region. 

Following the same procedure detailed in Section~\ref{subsec:single_layer}, the mode-matching equations Eqs. \eqref{eq:Ex_upper}-\eqref{eq:Ey_lower} are applied at $z=0$ and $z=-h$, coupling the outer FB harmonics $\mathcal{R}_{\mathbf{n}}$, $\mathcal{T}_{\mathbf{n}}$ to the waveguide modes $\mathcal{A}_{\pm,\mathbf{m}}^{(i),(1)}$ and $\mathcal{A}_{\pm,\mathbf{m}}^{(i),(2)}$, respectively (overall $4(N_x+1)(N_y+1)$ equations); and adapted versions of these electric-field equations are applied at $z=-h_1$ and $z=-h_1-\Delta h$, coupling the intermediate air-gap harmonics $\mathcal{M}_{\pm,\mathbf{n}}$ to the waveguide modes of both layers (overall another $4(N_x+1)(N_y+1)$ equations). Similarly, the magnetic field mode-matching equations Eqs. \eqref{eq:Hx_upper}-\eqref{eq:Hy_lower} are enforced at all four interfaces, projected onto the waveguide modal basis of the corresponding perforation set and layer index $l$ (adding $2(N_{p,1}+N_{p,2})(2M_xM_y+M_x+M_y)$ equations altogether). Combining all contributions, the boundary conditions at the four interfaces yield $U_{\mathrm{dbl}}=8(N_x+1)(N_y+1)+2(N_{p,1}+N_{p,2})(2M_xM_y+M_x+M_y)$ linear equations, assembled into a matrix system analogous to Eq. \eqref{eq:matrix_system}, with the unknown vector $\mathbf{I}_{U_{\mathrm{dbl}}\times 1}$ now additionally including the intermediate region FB coefficients $\mathcal{M}_{\pm,\mathbf{n}}^{\text{TM/TE}}$ and the waveguide modal amplitudes $\mathcal{A}_{\pm,\mathbf{m}}^{\text{TM},(i),(l)}$ and $\mathcal{A}_{\pm,\mathbf{m}}^{\text{TE},(i),(l)}$ for both layers. The resulting system can be cast in a block-matrix form analogous to Eq. \eqref{eq:matrix_system}, with the augmented $\mathbf{Z}_{U_{\mathrm{dbl}}\times U_{\mathrm{dbl}}}$, $\mathbf{I}_{U_{\mathrm{dbl}}\times 1}$, and $\mathbf{V}_{U_{\mathrm{dbl}}\times 1}$ structured similarly to their single-layer counterparts detailed in Appendix \ref{appendix:coefficients}, but expanded to accommodate the additional intermediate-region FB amplitudes $\mathcal{M}_{\pm,\mathbf{n}}^{\text{TM/TE}}$ and the waveguide modal amplitudes of both layers. 
Finally, as in the single-layer case, the resulting system may be readily solved via matrix inversion, $\mathbf{I} = \mathbf{Z}^{-1}\mathbf{V}$, yielding the complete set of scattered field amplitudes everywhere in space and -- importantly -- enabling subsequent MG design and optimization (see Sections \ref{subsec:synthesis} and \ref{subsec:te_double}).

\subsection{MG Synthesis Procedure}
\label{subsec:synthesis}
Once the scattering problem is solved via Eq. \eqref{eq:matrix_system} or its double-layer equivalent (Section \ref{subsec:double_layer}), the retrieved transmission coefficients $\mathcal{T}_{\mathbf{n}}^{\text{TE}}$ and $\mathcal{T}_{\mathbf{n}}^{\text{TM}}$ can be used to quantify the MG performance for a given geometry. Specifically, substituting this obtained solution into Eqs. \eqref{equ:incident_fields} and \eqref{equ:transmitted_fields} allows evaluation of the fraction of incident power coupled to the TE- or TM-polarized $(n_x,0)$-order transmitted FB harmonic, defined as the anomalous refraction efficiency for either polarization. For a given mode order $n_x$, the latter read
\begin{equation}
\begin{split}
\eta_{n_x,0}^{\text{TE}} &=
  \frac{|\mathcal{T}_{\mathbf{n}}^{\text{TE}}|^2\, k\, k_z^{(n_x)}}
       {|E_0^{\text{TE}}/\eta|^2 \cos\theta_{\text{in}}
        \left[k_x^{(n_x)}\right]^2}, \\[4pt]
\eta_{n_x,0}^{\text{TM}} &=
  \frac{|\mathcal{T}_{\mathbf{n}}^{\text{TM}}|^2\, k\, k_z^{(n_x)}}
       {|H_0^{\text{TM}}|^2 \cos\theta_{\text{in}}
        \left[k_x^{(n_x)}\right]^2},
\end{split}
\label{eq:efficiency}
\end{equation}
where $k_z^{(n_x)}=\sqrt{k^2-[k_x^{(n_x)}]^2}$ is the longitudinal wavenumber of the $(n_x,0)$-order transmitted mode. In the anomalous refraction MG case studies considered herein (Section \ref{sec:results}), our goal would be to maximize these efficiencies -- i.e., require that $\eta_{-1,0}^{\text{TE}}\to 1$ or $\eta_{-1,0}^{\text{TM}}\to 1$ (or both, depending on the target polarization response), 
which implies that coupling to all other propagating FB modes must be suppressed via proper tuning of the MG degrees of freedom.

Specifically, to synthesize an MG that implements the desired beam-deflection functionality, the geometric and constitutive parameters of the configuration -- lateral dimensions $d_x^{(i)}\times d_y^{(i)}$ and center coordinates $(a_x^{(i)}, a_y^{(i)})$ of the $i$th perforation, periodicity $L_y$, slab thickness $h$, number of perforations per unit cell $N_p^{(l)}$, dielectric loading $\varepsilon_{d,r}^{(i)}$, and, in the case of a double-layer configuration, individual layer thicknesses $h_1$, $h_2$ and air-gap height $\Delta h$ -- are adjusted as to approach the target efficiency values. This is accomplished by using the \textsc{Matlab} library function \texttt{fminsearch}\, utilizing repeated evaluations of Eq. \eqref{eq:efficiency} via the analytical model of Sections \ref{subsec:single_layer} and \ref{subsec:double_layer} to seek an optimum, yielding the desired MG layout.

For the double-layer configuration, an additional guided-wave suppression criterion is enforced during this procedure. Specifically, if resonant multiple reflections exist within the air gap (Fig. \ref{fig:double_layer}), the associated strong nonlocal effects -- which facilitate effective beam deflection within the periodic analysis -- may eventually lead to reduced performance for practical finite-area MGs \cite{Rabinovich2019}. To mitigate this, we minimize the power flowing laterally (along the $x$ axis) within the gap relative to the power $P_{\text{inc}}$ incident on the MG.

Formally, the lateral power $P_x(x_0)$ crossing the plane $x=x_0$ is obtained from the complex Poynting vector, $\mathbf{S}=\tfrac{1}{2}\mathbf{E}\times\mathbf{H}^*$, of the \emph{total} fields in the air gap, integrated over the cross section of the gap within one period, reading
\begin{equation}
P_x(x_0) = \frac{1}{2}\,\text{Re}\!\int_{0}^{L_y}\!\int_{-h_1-\Delta h}^{-h_1}\!\left[E_y H_z^* - E_z H_y^*\right]_{x=x_0}\,dz\,dy,
\label{eq:Px_poynting}
\end{equation}
with the fields calculated for the given MG configuration from Eqs. \eqref{equ:incident_fields}-\eqref{equ:transmitted_fields} and \eqref{equ:middle_fb}. Since cross-polarization coupling to guided and evanescent modes can take place in the air gap, it is crucial to take into account in the calculation the contribution of both TE- and TM-polarized FB harmonics therein (corresponding to $\mathcal{M}_{\pm,\mathbf{n}}^{\text{TE}}$ and $\mathcal{M}_{\pm,\mathbf{n}}^{\text{TM}}$). We further note that the dependence of the lateral power on the plane of reference $x_0$ stems from the cross-modal terms of the Poynting vector, with periodicity coercing $P_x(x_0)=P_x(x_0+L_x)$. Hence, to control this power-flow component globally, we consider its maximum value within the unit cell for synthesis purposes. 

In other words, the synthesis seeks
\begin{equation}
\min\left[\frac{\max\limits_{{0\leq}x_0<L_x}\left|P_x(x_0)\right|}{P_{\text{inc}}}\right],
\label{eq:guided_wave_condition}
\end{equation}
quantifying the extent of nonlocal power flow required to sustain the anomalous refraction in our design. In Eq. \eqref{eq:guided_wave_condition}, the overall incident power $P_\mathrm{inc}$ illuminating the relevant unit cell is evaluated by integrating the $z$-component of the same Poynting vector defined above for the incident fields of Eq. \eqref{equ:incident_fields} over a single period,
yielding $P_{\text{inc}}^{{\mathrm{TE}}}=\tfrac{1}{2}\left(|E_0^{\text{TE}}|^2/\eta\right)\cos\theta_{\text{in}}L_xL_y$ and $P_{\text{inc}}^{{\mathrm{TM}}}=\tfrac{1}{2}\eta|H_0^{\text{TM}}|^2\cos\theta_{\text{in}}L_xL_y$ for TE- and TM-polarized excitations, respectively. 

By employing these steps (for either the single- or double-layer configurations), the analytical model is leveraged to solve the inverse design problem, directly yielding -- as shall be demonstrated in Section \ref{sec:results} to follow -- fabrication-ready MG specifications for realizing the prescribed dual-polarized anomalous refraction without resorting to full-wave optimization. 

\section{Results and Discussion}
\label{sec:results}
To demonstrate the versatility and accuracy of the proposed analytical framework, we follow the synthesis procedure outlined in Section \ref{subsec:synthesis} to devise and validate four MG designs of increasing complexity. In all designs, the MGs are illuminated from the upper half-space ($z>0$) by a plane wave incident at $\theta_{\text{in}}=20\degree$, and are engineered to deflect the incident power towards the anomalous refraction angle $\theta_{\text{out}}=-50\degree$ at an operating frequency of $20\,\mathrm{GHz}$ (free-space wavelength $\lambda\approx15\,\mathrm{mm}$). The periodicity along $x$ is accordingly set to $L_x=\lambda/|\sin\theta_{\text{in}}-\sin\theta_{\text{out}}|\approx 13.53\,\mathrm{mm}$, while $L_y<\lambda$ confines radiative scattering to the $\widehat{xz}$ plane \footnote{As discussed in Section \ref{subsec:synthesis}, $L_y$ is considered a degree of freedom within the synthesis framework. Thus, in each of the following subsections (Sections \ref{subsec:tm_single} to \ref{subsec:dual}), the optimal value of $L_y$ (within the constraint $L_y<\lambda$) chosen for the respective MG will be specified.}, ensuring that only two FB modes propagate: the specular $(0,0)$ order and the anomalous $(-1,0)$ order.

\subsection{Single-Layer All-Metallic MG for\\TM-Polarized Anomalous Refraction}
\label{subsec:tm_single}
\begin{figure}[!t]
\centerline{\includegraphics[width=1.0\columnwidth]{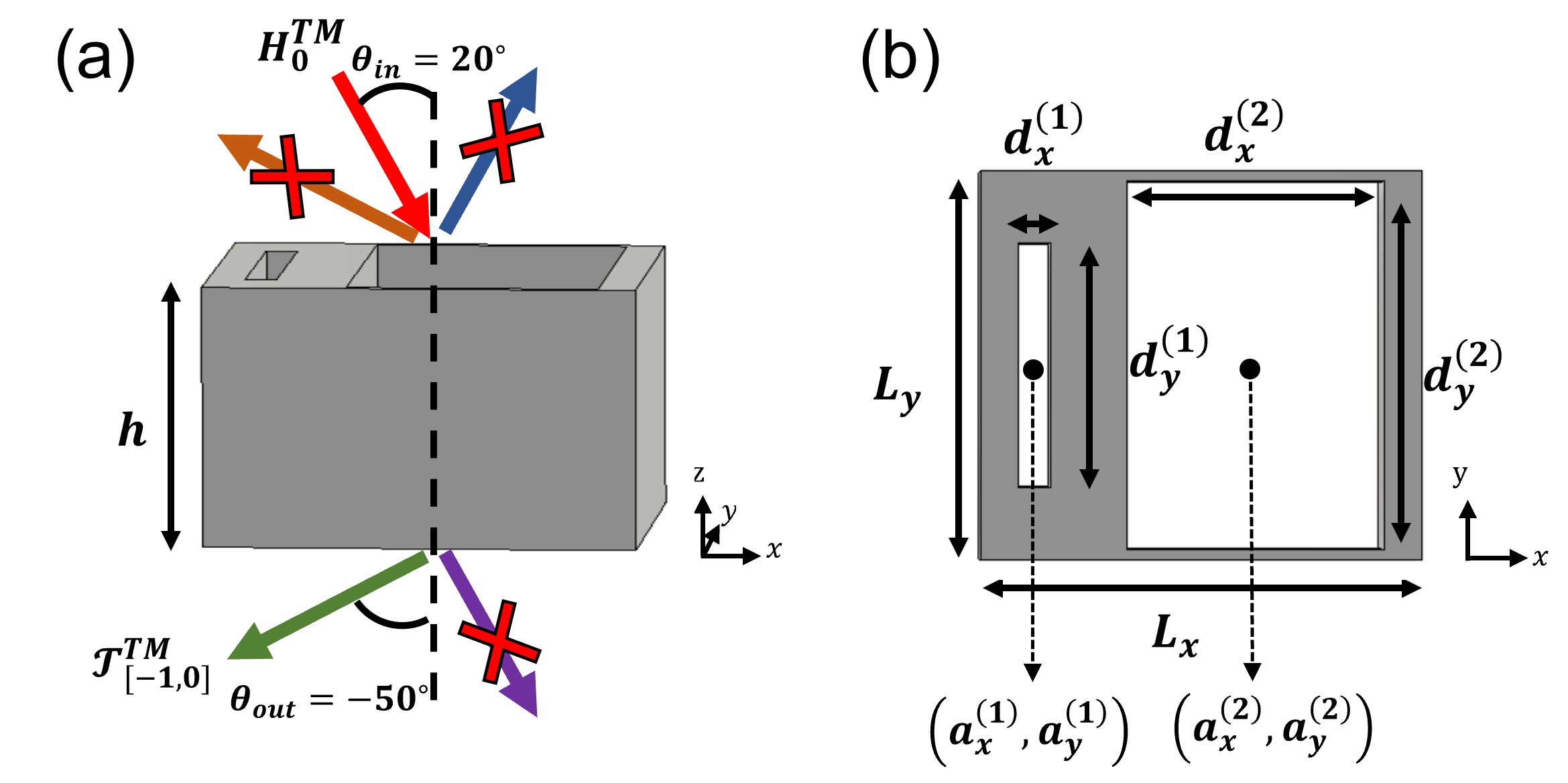}}
\caption{Single-layer all-metallic MG for TM-polarized anomalous refraction from $\theta_{\text{in}}=20\degree$ towards $\theta_{\text{out}}=-50\degree$ at $f=20\,\mathrm{GHz}$ (Section \ref{subsec:tm_single}), featuring four propagating FB modes. (a) Three-dimensional scattering scenario; the incident TM-polarized beam is anomalously refracted towards $\theta_{\text{out}}$, while spurious propagating modes are suppressed (marked with $\times$). (b) Top view of the unit cell ($N_p=2$ perforations per period), with the relevant geometric parameters indicated.}
\label{fig:design1_config}
\end{figure}

For the first demonstration of the proposed framework, we choose to devise a metallic MG for realization of TM-polarized anomalous refraction. Considering the natural support of TM-polarized wave manipulation exhibited by metallic MG constructs observed in the past \cite{Rabinovich2020_1, Rahmanzadeh2020, Hemmatyar2021, Rajabalipanah2022, Rahmanzadeh2022}, we expect this to be feasible with minimal complexity. In particular, we approach this target functionality with a rather simple all-metallic ($\varepsilon_{d,r}=1$) single-layer configuration; to suppress both forward and backward spurious FB modes [three in total, \textit{cf.} Fig. \ref{fig:geometry}(b)], we replace the single-groove solution used for the single-polarized anomalous \emph{reflection} MG (one spurious FB harmonic) considered in \cite{Rabinovich2020_1} with an extended $N_p=2$ perforations-per-period structure for the transmissive case of interest herein [Fig. \ref{fig:design1_config}(a)]. 

Following the procedure outlined in Section \ref{subsec:synthesis}, we apply the MATLAB library function \texttt{fminsearch} to optimize the geometric degrees of freedom of the perforated metallic  slab using the analytical model of Section \ref{subsec:single_layer}, seeking the configuration that maximizes $\eta_{-1,0}^{\text{TM}}$ at $f=20\,\mathrm{GHz}$. This yields a slab of thickness $h = 11\,\mathrm{mm}$, with the perforations' lateral dimensions and center coordinates given by $d_x^{(1)} = 1.0\,\mathrm{mm}$, $d_y^{(1)} = 7.5\,\mathrm{mm}$, and $(a_x^{(1)}, a_y^{(1)}) = (-5.1\,\mathrm{mm}, 0)$; and $d_x^{(2)} = 7.87\,\mathrm{mm}$, $d_y^{(2)} = 11.28\,\mathrm{mm}$, and $(a_x^{(2)}, a_y^{(2)}) = (1.66\,\mathrm{mm}, 0)$; for the $i=1$ and $i=2$ perforations, respectively; the optimized periodicity along $y$ is $L_y = 11.9\,\mathrm{mm}$ [Fig. \ref{fig:design1_config}(b)]. 

\begin{figure}[!t]
\centerline{\includegraphics[width=1.0\columnwidth]{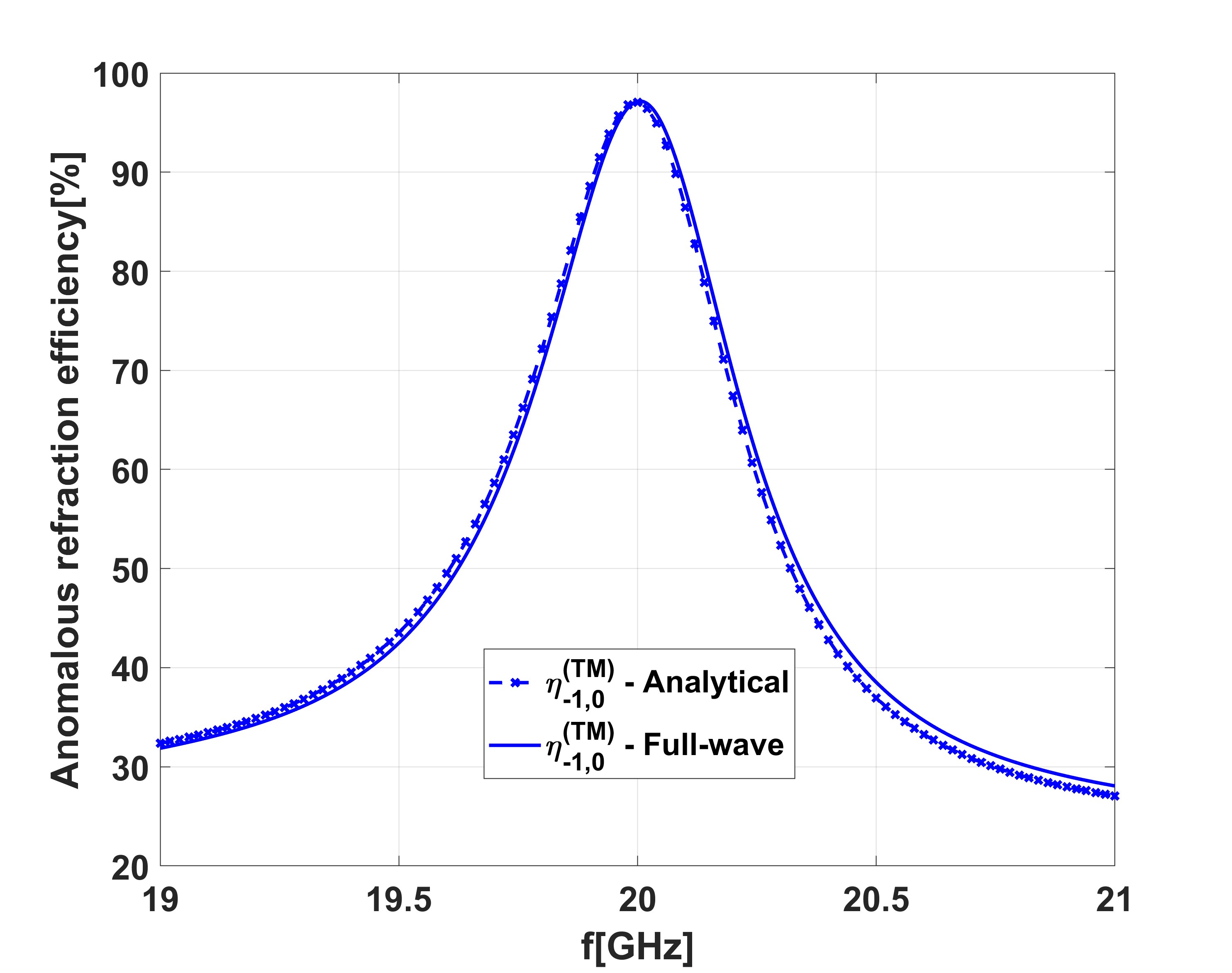}}
\caption{Anomalous refraction efficiency $\eta_{-1,0}^{\text{TM}}$ for the single-layer all-metallic MG (Section \ref{subsec:tm_single}). Analytical mode-matching predictions (blue dashed line with markers) and full-wave simulation results (blue solid line) show excellent agreement, with a peak efficiency of $97\%$ into the $(-1,0)$-order mode at $20\,\mathrm{GHz}$.}
\label{fig:design1_efficiency}
\end{figure}

\begin{figure}[!t]
\centerline{\includegraphics[width=0.8\columnwidth]{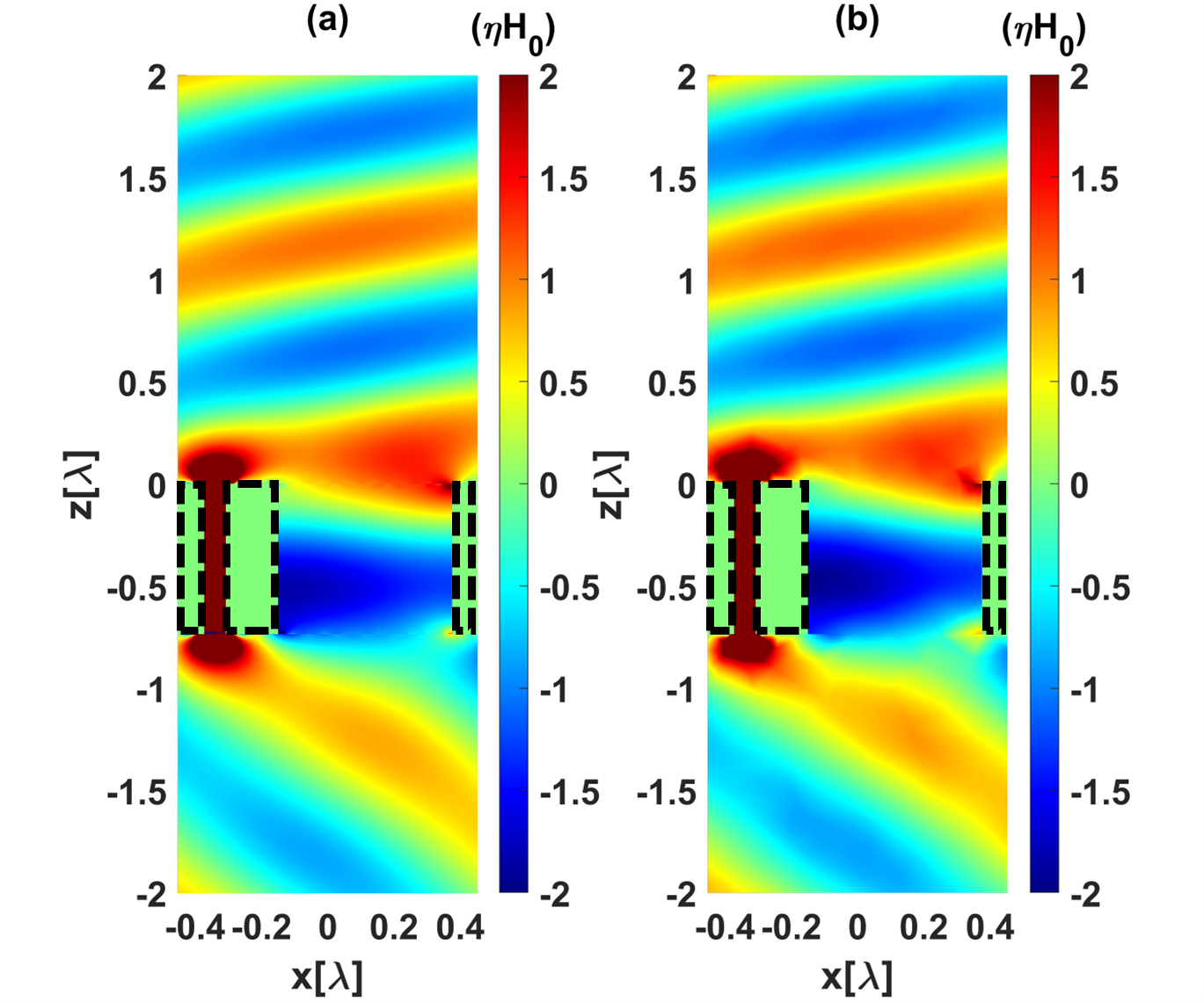}}
\caption{Spatial distribution of $\text{Re}\{E_x(x,y=0,z)\}$ for the single-layer all-metallic MG at $20\,\mathrm{GHz}$. Comparison between (a) analytical mode-matching predictions and (b) full-wave simulation results shows excellent agreement in both field amplitude and phase distribution.}
\label{fig:TM_fields}
\end{figure}

Figure \ref{fig:design1_efficiency}  presents the evaluated $\eta_{-1,0}^{\text{TM}}$ for this prototype design as a function of frequency, comparing analytical mode-matching predictions (Section \ref{sec:theory}) with the results recorded in full-wave simulations performed using CST Microwave Studio using the obtained parameters. Excellent agreement is observed across the entire frequency range, with both approaches predicting a peak efficiency exceeding $97\%$ at the design frequency of $f=20$ GHz, validating the accuracy of the analytical framework for this configuration. The field snapshots across the plane of incidence $\text{Re}\{E_x(x,y=0,z)\}$ displayed in Fig. \ref{fig:TM_fields} provide further evidence for this agreement, showcasing remarkable correspondence between analytical [Fig. \ref{fig:TM_fields}(a)] and full-wave [Fig. \ref{fig:TM_fields}(b)] results, clearly visualizing the coupling from the incident plane wave at $\theta_{\text{in}} = 20\degree$ (upper region, $z>0$), through the truncated waveguides ($-h < z < 0$), to the anomalously refracted wave at $\theta_{\text{out}} = -50\degree$ (lower region, $z<-h$).

Beyond validating the fundamental capability of the proposed framework to design simple all-metallic single-layer MGs for efficient TM-polarized anomalous refraction, these results confirm the high fidelity of the underlying mode-matching formalism, thereby establishing a solid baseline for the more complex designs that follow.

\subsection{Double-Layer All-Metallic MG for\\*TE-Polarized Anomalous Refraction}
\label{subsec:te_double}
As discussed in \cite{Rabinovich2020_1}, achieving efficient beam deflection for TE-polarized fields in all-metallic structures is inherently more challenging than for their TM-polarized counterparts. This may be attributed to the fundamentally different surface waveguiding properties exhibited by metallic interfaces for TE- and TM-polarized fields \cite{Collin1991}, and the critical role that coupling to this evanescent wave spectrum plays in facilitating effective diffraction engineering functionalities \cite{Rabinovich2020_1,Epstein2016_4,Epstein2017,DiazRubio2017,Rabinovich2020}. In the case of the transmissive anomalous refraction functionality considered herein, however, additional mechanisms may come into play. 

As noted in Section \ref{subsec:tm_single}, to facilitate sufficient degrees of freedom for suppressing three spurious FB modes (simplest single-polarized anomalous refraction operation), we utilize two perforations per period (e.g., \textit{cf.} Fig. \ref{fig:design1_config}). Since the plane of incidence is $\widehat{xz}$ and the beam deflection is along the $x$ axis -- requiring phase modulation along the same direction -- it makes sense to offset the centers of these two perforations accordingly, to predominantly perturb the slab profile along $x$ as well (again, as demonstrated in Fig. \ref{fig:design1_config}). Nonetheless, when such a configuration is used for wide-angle refraction ${|\sin\theta_\mathrm{in}-\sin\theta_\mathrm{out}|>1}$, as considered herein (and often for MGs in general \cite{Radi2017, Wong2018, Rabinovich2018, Rabinovich2020, Popov2020}), at least one of these perforations must have its edge length $d_x^{(i)}$ along the $x$-direction smaller than $\lambda/2$ (since ${\Lambda=\lambda/|\sin\theta_\mathrm{in}-\sin\theta_\mathrm{out}|<\lambda}$). In this perforation, \emph{all} $\mathrm{TM}_{m_x,m_y}$ \textit{waveguide} modes are \emph{below} cutoff (with the lowest order being $\mathrm{TM}_\mathrm{1,1}$), and only the fundamental $\mathrm{TE}_{m_x,m_y}=\mathrm{TE}_\mathrm{0,1}$ \textit{waveguide} mode will be propagating \cite{Pozar2012}. Since the effective degrees of freedom in such waveguide-based MGs are tightly related to the total number of propagating modes within a period \cite{Dong2020}, the latter observation means that only incident plane waves that can couple effectively to the $\mathrm{TE}_\mathrm{0,1}$ waveguide modes -- for which $H_x=E_y=E_z=0$ -- will be able to exploit the full potential of the (essential) two-perforation setup for diffraction engineering \footnote{Indeed, to effectively tune the wavefront at $z<0$, significant phase accumulation should be achieved while the waves are transported through the perforations, which is impractical for below-cutoff (and even near-cutoff) modes. As representative case studies, one may consider the wavelength-comparable thicknesses $h$ of the optimized prototypes presented in Sections \ref{subsec:tm_single}, \ref{subsec:pbs}, and \ref{subsec:dual} (accounting for the effective wavelength when the perforation is loaded by a dielectric), which would lead to substantial decay of below-cutoff modes and correspondingly-ineffective power transmission.}. As a direct consequence, such MG configurations cannot efficiently manipulate TE-polarized waves impinging from $z>0$, since their only non-vanishing tangential components are $E_y$ and $H_x$ [\textit{cf.} Eq. \eqref{equ:incident_fields}], completely incompatible (in terms of continuity conditions) with the ones supported with the $\mathrm{TE}_\mathrm{0,1}$ mode. 

Thus, as a means for overcoming this obstacle, we introduce an additional metallic (perforated) layer and an intermediate air gap (\textit{cf.} Fig. \ref{fig:double_layer}), increasing the number of degrees of freedom with respect to the simpler single-layer TM-polarized design of Section \ref{subsec:tm_single} in an attempt to obtain high-efficiency TE-polarized anomalous refraction through this augmented design space -- still within the frame of an all-metallic MG construct. Beyond the larger overall effective degrees of freedom, the introduction of the intermediate air gap itself provides another physical mechanism to manipulate the overall scattering, allowing control over the interlayer coupling through multiple reflection phenomena. Importantly, this mechanism is partially accessible also to below-cutoff modes, which can tunnel through the waveguides and benefit from substantial phase accumulation in the multimodal Fabry-Perot cavity formed between the two perforated slabs. 

\begin{figure}[!t]
\centerline{\includegraphics[width=\columnwidth]{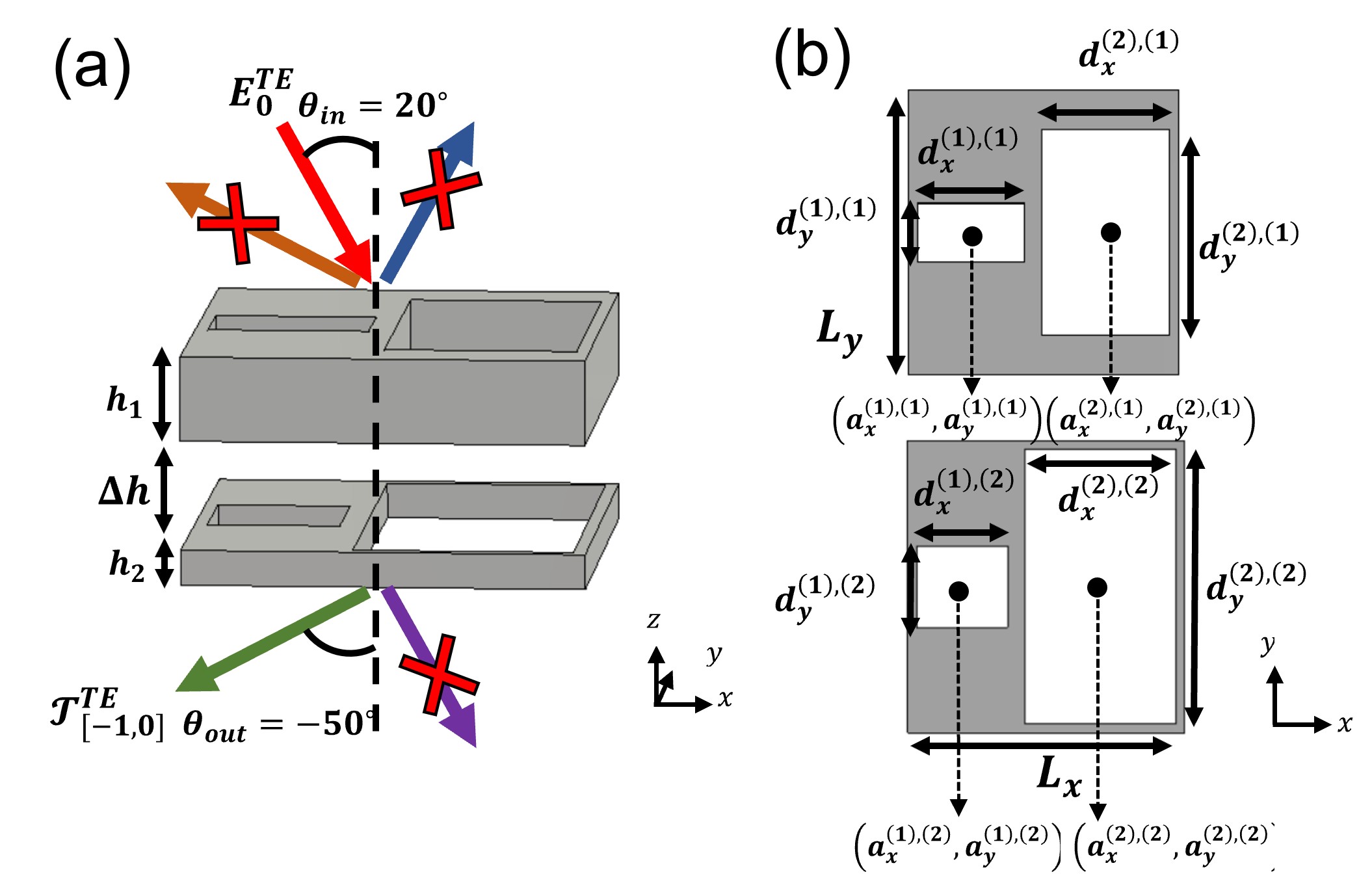}}
\caption{Double-layer all-metallic MG for TE-polarized anomalous refraction from $\theta_{\text{in}}=20\degree$ towards $\theta_{\text{out}}=-50\degree$ at $f=20\,\mathrm{GHz}$ (Section \ref{subsec:te_double}). (a) Three-dimensional scattering scenario; the incident TE-polarized beam is anomalously refracted towards $\theta_{\text{out}}$, while spurious propagating modes are suppressed (marked with $\times$). (b) Top views of the unit cells of the first ($l=1$, top) and second ($l=2$, bottom) layers ($N_p=2$ perforations per period each), with the relevant geometric parameters indicated.}

\label{fig:design2_config}
\end{figure}
Armed with this expanded scope for exploration, we proceed with the desired MG synthesis by following once again the procedure of Section \ref{subsec:synthesis}, adjusted now to address the double-layer configuration based on the formulation of Section \ref{subsec:double_layer}. Our goal is to determine the MG parameter specifications [corresponding to the configuration in Fig. \ref{fig:design2_config}(a)] that maximize $\eta_{-1,0}^{\text{TE}}$ at $f=20\,\mathrm{GHz}$. Also for this task, we choose to utilize $N_{p,1} = N_{p,2} = 2$ rectangular perforations per unit cell in each layer, balancing the number of independent waveguides within the period -- providing opportunities to tune modal phase accumulation and mode-matching-induced scattering coefficients -- with the wish to retain adequate edge lengths along $x$ ($d_x^{(i)}$) as to limit the decay rate (enhance power transmission) of below-cutoff waveguide modes.

Correspondingly, similar to Section \ref{subsec:tm_single}, we apply the library function \texttt{fminsearch} to optimize the geometric degrees of freedom via the analytical model of Section \ref{subsec:double_layer}. Note that, as discussed in Section \ref{subsec:synthesis}, in this case we further invoke the criterion of Eq. \eqref{eq:guided_wave_condition} to reduce undesirable excitation of guided power between the top and bottom slabs. Executing thus the synthesis procedure yields, as the lowest-ratio solution found, a double-layer MG design (Fig. \ref{fig:design2_config}) having an air gap of height $\Delta h = 3.68\,\mathrm{mm}$ separating the top and bottom layers, featuring thicknesses of $h_1 = 2.85\,\mathrm{mm}$ and $h_2 = 1.17\,\mathrm{mm}$, respectively, for a total MG thickness of $h = h_1+\Delta h+h_2 = 7.7\,\mathrm{mm}$. The two perforations in the top layer ($l=1$) feature lateral dimensions $d_x^{(1,1)} \times d_y^{(1,1)} = 5.39\,\mathrm{mm}\times 2.93\,\mathrm{mm}$, $d_x^{(2,1)}\times d_y^{(2,1)} = 6.43\,\mathrm{mm}\times 10.35\,\mathrm{mm}$ and center coordinates $(a_x^{(1,1)}, a_y^{(1,1)}) = (-3.64\,\mathrm{mm}, 0)$ and $(a_x^{(2,1)}, a_y^{(2,1)}) = (3.12\,\mathrm{mm}, 0)$. The second layer features $d_x^{(1,2)} = 4.45\,\mathrm{mm}$, $d_y^{(1,2)} = 3.98\,\mathrm{mm}$, center coordinates $(a_x^{(1,2)}, a_y^{(1,2)}) = (-4.11\,\mathrm{mm}, 0)$, and $d_x^{(2,2)} = 7.33\,\mathrm{mm}$, $d_y^{(2,2)} = 13.44\,\mathrm{mm}$, center coordinates $(a_x^{(2,2)}, a_y^{(2,2)}) = (2.65\,\mathrm{mm}, 0)$. Both layers share the same periodicity along $y$, optimized as $L_y^{(1)}=L_y^{(2)} = 14.3\,\mathrm{mm}$.

\begin{figure}[!t]
\centerline{\includegraphics[width=1.0\columnwidth]{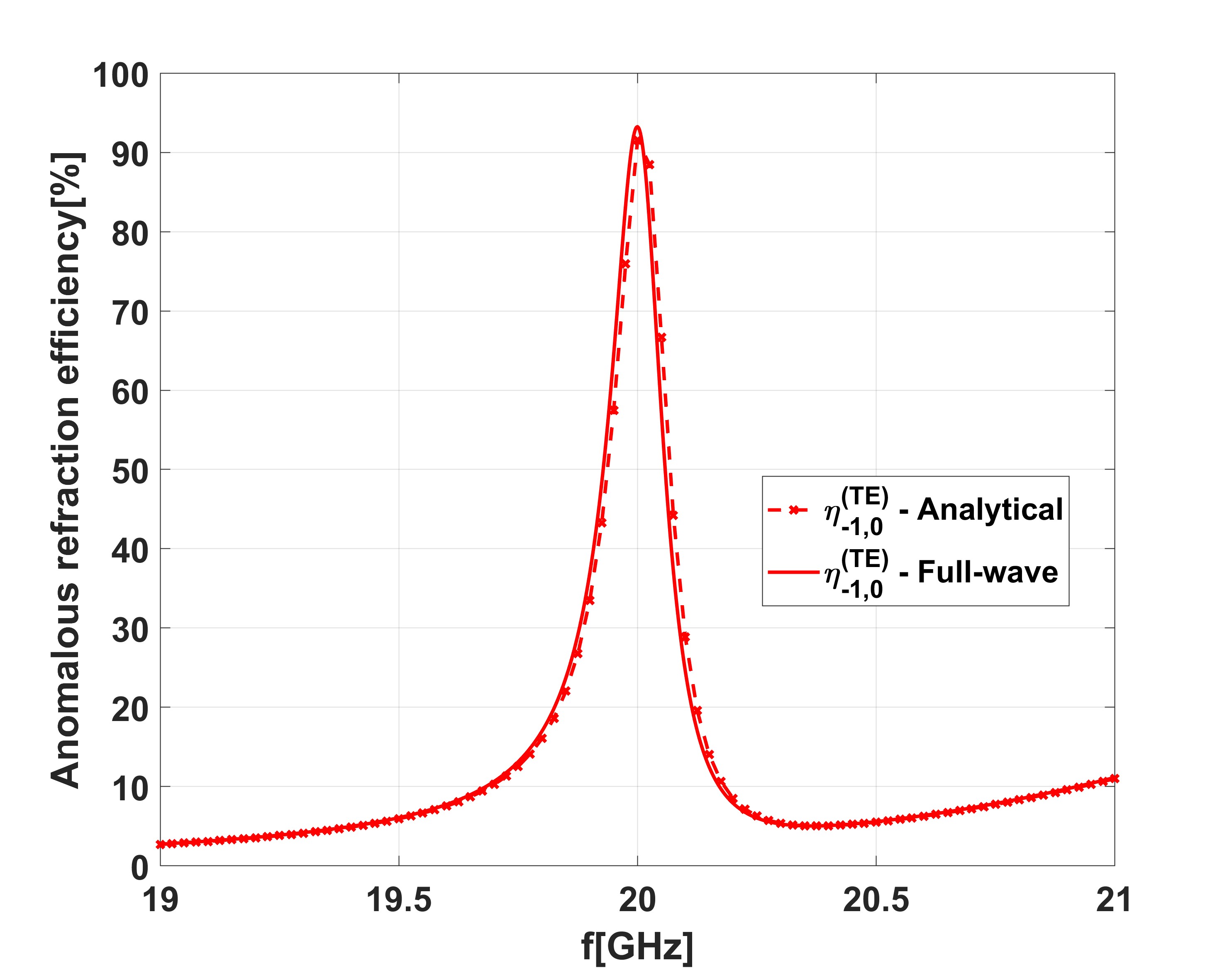}}
\caption{TE-polarized anomalous refraction efficiency $\eta_{-1,0}^{\text{TE}}$ of the double-layer all-metallic MG (Section \ref{subsec:te_double}) as a function of frequency. Analytical predictions (dashed line) are compared with full-wave simulation results (solid line).}
\label{fig:design2_efficiency}
\end{figure}

To validate the obtained design, we define the double-layer configuration in CST Microwave Studio and record the simulated scattered fields when the MG is illuminated by the designated incident plane wave. Figure \ref{fig:design2_efficiency} presents the corresponding anomalous refraction efficiency $\eta_{-1,0}^{\text{TE}}$ as a function of frequency, comparing the analytical mode-matching predictions with the full-wave simulation results.

As can be clearly seen, the two curves agree very well, revealing that the optimal operating point indeed coincides with the design frequency $20\,\mathrm{GHz}$, with the coupling efficiency peaking at approximately $92\%$ (the rest mostly contributes to direct transmission, with $\eta_{0,0}^{\text{TE}}\approx 7\%$). 

\begin{figure}[!t]
\centerline{\includegraphics[width=0.8\columnwidth]{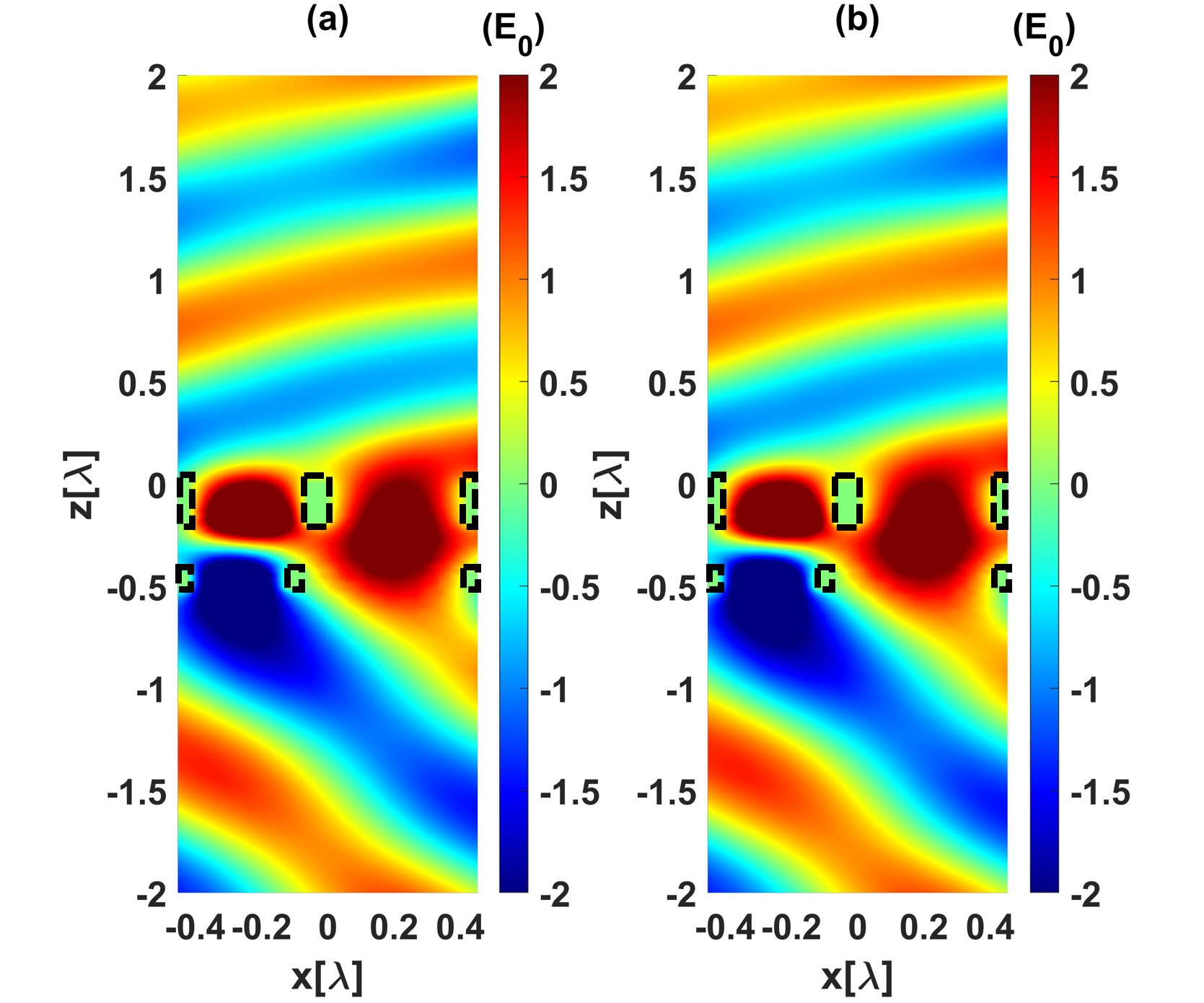}}
\caption{Electric field distribution $\text{Re}\{E_y(x,y=0,z)\}$ for the double-layer all-metallic MG (Section \ref{subsec:te_double}), when illuminated from $\theta_{\text{in}}=20\degree$ by a TE-polarized plane wave at $f=20\,\mathrm{GHz}$ (single period shown). (a) Analytical prediction and (b) full-wave simulation result. Black dashed lines mark the metallic slab boundaries.}
\label{fig:TE_fields}
\end{figure}

While this value is somewhat lower than the one obtained for TM-polarized beam deflection efficacy demonstrated in Section \ref{subsec:tm_single} -- as expected due to the challenging polarization \cite{Rabinovich2020_1} -- these results confirm that the augmented set of design levers offered by the devised double-layer architecture successfully enables high-efficiency TE-polarized anomalous refraction in an all-metallic configuration. As additional evidence to this observation, we compare in Fig. \ref{fig:TE_fields} the spatial distribution of the relevant electric field component in the $\widehat{xz}$ plane at $f=20\,\mathrm{GHz}$ as predicted analytically [Fig. \ref{fig:TE_fields}(a)] with the one obtained from full-wave simulations [Fig. \ref{fig:TE_fields}(b)]. The two field maps are in close agreement across all relevant regions of space, corroborating the accuracy of the mode-matching analysis while highlighting the anomalous refraction efficacy.

Indeed, these results indicate that the devised double-layer all-metallic architecture provides sufficient degrees of freedom for manipulating TE-polarized waves in transmission. Nonetheless, they also expose several practical limitations inherent to this topology. First, as already mentioned in the beginning of this subsection, transmission in the TE-polarized case inevitably relies on tunnelling of evanescent waveguide modes through the perforations. Consequently, since differential phase accumulation in the two lateral apertures cannot be exploited to realize the phase gradients required for beam steering, the desired response must primarily be achieved through multiple reflections within the slabs. This seemingly drives the design toward extreme disparities between the $y$-dimensions of the two perforations in each slab, corresponding to the dominant electric-field component of the incident wave. In practice, the implication is very large values of $d_y^{(2,1)}$ and especially $d_y^{(2,2)}$, which require correspondingly large values of $L_y$. These near-wavelength ultimate dimensions for $L_y$, in turn, introduce a practical risk: if the illuminating wave has even a minimal angular spread in the $y$ direction -- inevitable in practice -- additional $n_y \neq 0$ FB modes (grating lobes, in essence) may become weakly excited, changing the partition between radiation channels and reducing the overall coupling to the desired anomalous refraction mode.

Second, as we have pointed out in Section \ref{subsec:synthesis} surrounding Eq. \eqref{eq:guided_wave_condition}, the same reliance on multiple reflections for obtaining the target functionality may give rise to substantial power being guided laterally inside the air gap. Such confined modes propagating along the $x$ (or $y$) axis between the slabs -- very challenging to suppress in practice -- may prove to be detrimental in the ultimate finite-size implementation, which cannot provide sufficient means for their build-up at the device periphery. However, without such build-up, the nonlocal field modulation supporting the anomalous refraction would not be properly established \cite{Epstein2016_4,Tcvetkova2018}, violating the boundary conditions used to ensure effective beam deflection  for the periodic configuration in the design stage.

Lastly, as can be clearly seen in Fig. \ref{fig:design2_efficiency}, the same multiple reflection phenomenon forms a resonant construct, inherently limiting the operational bandwidth. This effect may be even more pronounced when coupled with the limited spatial spectral bandwidth mentioned in relation to the large $L_y$ values, potentially restricting applicability in wideband scenarios. 

Taken together, these observations suggest that it may be worthwhile to consider alternative approaches for achieving effective TE-polarized field manipulation, which avoid metallic layer cascades prone to these realization challenges. In the subsequent Sections \ref{subsec:pbs} and \ref{subsec:dual} we present such an approach, utilizing dielectric loading in order to demonstrate high-efficiency dual-polarized anomalous refraction.

\subsection{Single-Layer Dielectric-Loaded MG for Polarization Beam Splitting}
\label{subsec:pbs}
In view of the challenges highlighted towards the end of the previous section with respect to application of MG cascades in practical scenarios, we wish to explore alternative paths to introduce degrees of freedom to the basic single-layer metallic configuration of Section \ref{subsec:tm_single}, as to facilitate efficient manipulation of TE-polarized fields as well. As we recall from our discussions in Sections \ref{sec:introduction} and \ref{subsec:te_double}, the challenge is related to the dominant coupling of TE-polarized field excitation to below-cutoff modes within the perforations. In general, as observed in \cite{Dong2020}, the effective degrees of freedom available for diffraction engineering are tightly related to the overall number of propagating modes in the $N_p$ waveguides spanning a single period of the MG. One way to increase this number while still retaining a single-layer configuration is by utilizing dielectric loading within the metallic perforations: introduction of a dielectric of relative permittivity $\varepsilon_{d,r}$ would reduce the effective wavelength (and thus the cutoff frequency) within the perforation by a factor of $\sqrt{\varepsilon_{d,r}}$, thereby allowing more modes to propagate at a given operating frequency -- despite the constraints on the lateral dimensions (waveguide cross-sections) imposed by the target functionality and the MG period (see discussion in Section \ref{subsec:te_double}).

\begin{figure}[!t]
\centerline{\includegraphics[width=1.0\columnwidth]{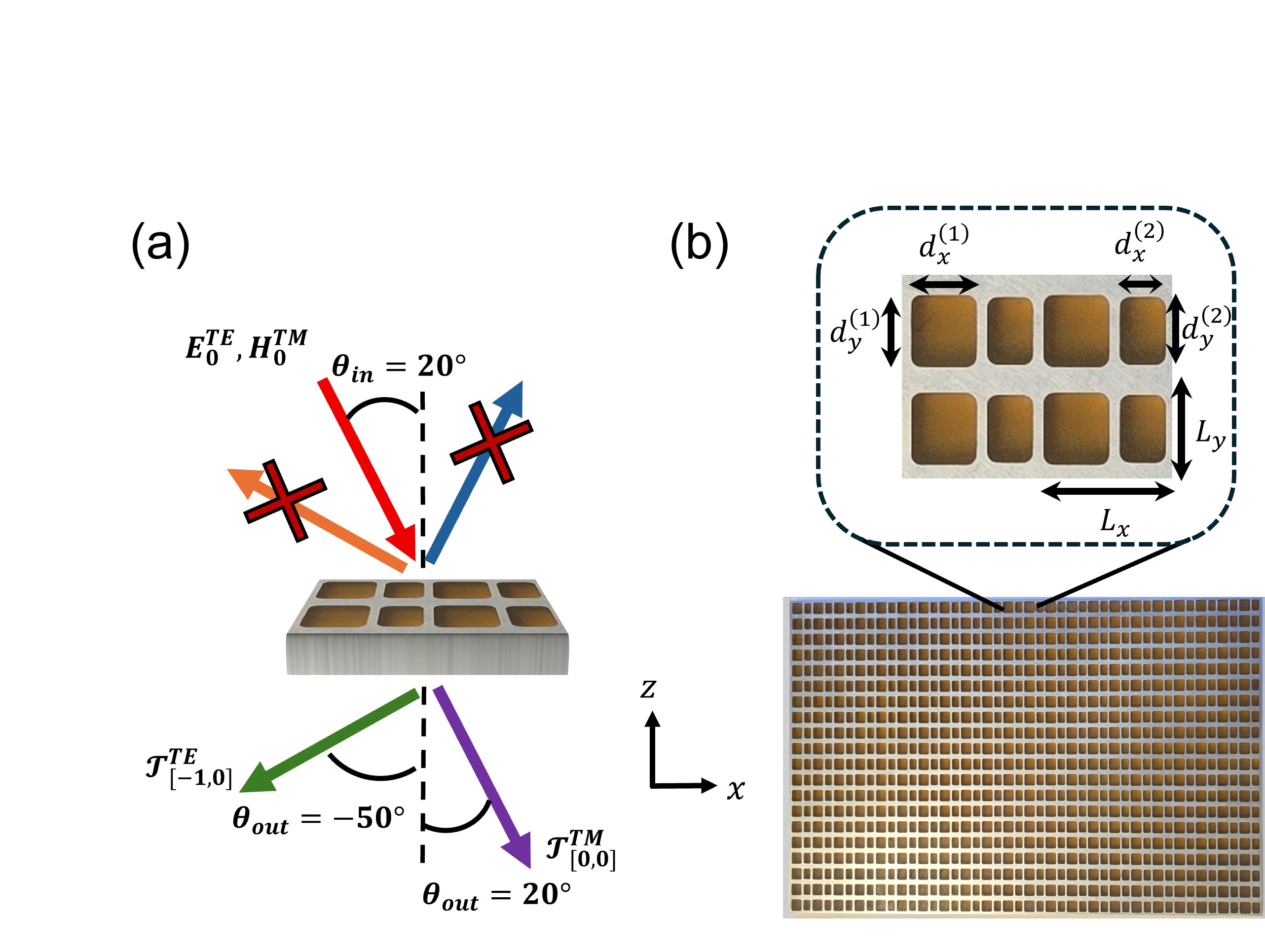}}
\caption{Dielectric-loaded MG for polarization beam splitting (Section \ref{subsec:pbs}). (a) Three-dimensional scattering scenario, illustrating the target dual-polarized beam-splitting functionality: the TE-polarized incident field is anomalously refracted towards $\theta_{\text{out}}=-50^\circ$, while the TM-polarized incident field is transmitted directly towards $\theta_{\text{out}}=20^\circ$. (b) Full view of the fabricated prototype: CNC-machined aluminum panel with Ultem 1000 dielectric inserts; inset: closeup on the unit cell, with the relevant geometric parameters indicated.}
\label{fig:TE1TM0_fabricated}
\end{figure}

Indeed, we proceed herein accordingly and employ this approach to further extend the effectiveness of TE-polarized beam manipulation, aiming at achieving dual-polarized diffraction control in such single-layer dielectric loaded MGs. Specifically, we fill the perforations within the metallic MG frame with Ultem 1000, a solid dielectric insert characterized by a dielectric constant of $\varepsilon_{d,r} = 3.0$ and loss tangent $\tan\delta = 0.002$. Although this loading strategy partially sacrifices some of the advantages associated with the strictly all-metallic structures discussed in Section \ref{sec:introduction}, Ultem 1000's high glass transition temperature ($T_g=217\,^\circ$C) and machinability, demonstrated by our utilized fabrication process described below, preserve many of the practical benefits of the original architecture, while simultaneously paving the way for the deployment of MGs in environmentally demanding applications requiring dual-polarized operation. 

To demonstrate the latter, we aim at devising in this subsection a polarization beam splitting MG, whose target functionality is illustrated in Fig.~\ref{fig:TE1TM0_fabricated}(a): an unpolarized plane wave at $f=20$~GHz impinging from $\theta_{\text{in}}=20^\circ$ will be separated into its polarization components upon transmission, with the TM-polarized fields continuing to propagate uninterrupted through the MG towards $\theta_{\text{out}}=20^\circ$, and the TE-polarized portion undergoing anomalous refraction towards $\theta_{\text{out}}=-50^\circ$. Correspondingly, to address this dual-polarized design task, the synthesis procedure of Section \ref{subsec:synthesis} is applied to simultaneously satisfy both the TE and TM efficiency conditions $\eta_{0,0}^{\text{TM}} \to 1$ and $\eta_{-1,0}^{\text{TE}} \to 1$. Being able now to manipulate effectively both TM- and TE- polarized fields due to the augmented configuration, we attempt to achieve this using $N_p = 2$ rectangular Ultem-filled perforations per period, balancing the more challenging requirement for fulfilling two efficiency constraints simultaneously with the extended number of degrees of freedom (increased number of propagating waveguide modes) available due to the dielectric loading.

\begin{figure}[!t]
\centerline{\includegraphics[width=1.0\columnwidth]{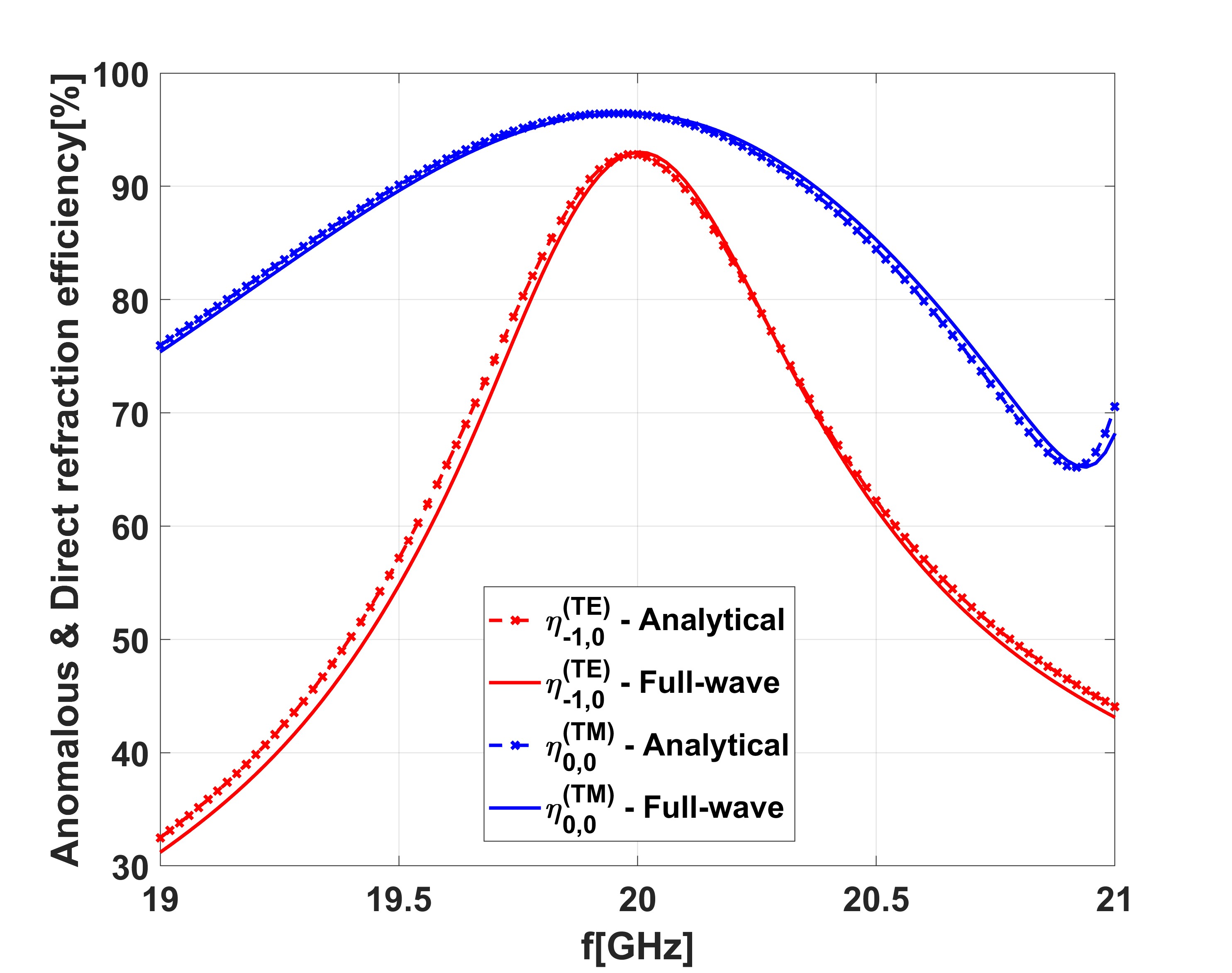}}
\caption{Anomalous refraction efficiencies of the dielectric-loaded polarization beam splitter (Section \ref{subsec:pbs}) as a function of frequency, quantifying the relative co-polarized power coupled to the $(-1,0)$ mode for TE-polarized incident fields (red) and to the fundamental $(0,0)$ mode for TM-polarized excitations (blue), as designated. Analytical predictions (dashed lines) are compared with full-wave simulation results (solid lines).}
\label{fig:design3_analytical}
\end{figure}

Invoking thus the scheme outlined in Section \ref{subsec:synthesis} in conjunction with the analytical model of Section \ref{subsec:single_layer} yields the design specifications of the dielectric-loaded MG, featuring a perforated metallic slab of thickness $h = 6.3\,\mathrm{mm}$, with the perforations' lateral dimensions and center coordinates given by $d_x^{(1)} = 4.64\,\mathrm{mm}$, $d_y^{(1)} = 7.0\,\mathrm{mm}$, and $(a_x^{(1)}, a_y^{(1)}) = (-3.9\,\mathrm{mm}, 0)$ for the first perforation ($i=1$); and $d_x^{(2)} = 6.9\,\mathrm{mm}$, $d_y^{(2)} = 7.3\,\mathrm{mm}$, and $(a_x^{(2)}, a_y^{(2)}) = (2.8\,\mathrm{mm}, 0)$ for the second perforation ($i=2$); the optimized periodicity along $y$ is $L_y = 9.9\,\mathrm{mm}$ [Fig. \ref{fig:TE1TM0_fabricated}(b)]. The expected performance of this design is assessed in Fig. \ref{fig:design3_analytical}, which compares the analytically evaluated (dashed lines with markers) and full-wave simulated (solid lines) transmission efficiencies for both polarizations as a function of frequency. The two sets of curves are in excellent agreement, predicting successful coupling of TE-polarized incident fields to the anomalous $(-1,0)$-order mode (red) and direct transmission (dominant coupling to the $(0,0)$-order mode) of TM-polarized waves (blue). Both coupling efficiencies peak at the designated operating frequency $f=20\,\mathrm{GHz}$, with maximal values of $93\%$ and $97\%$, respectively, thereby validating the analytical model for this dielectric-loaded dual-polarized scenario.

\begin{figure}[!t]
\centerline{\includegraphics[width=0.8\columnwidth]{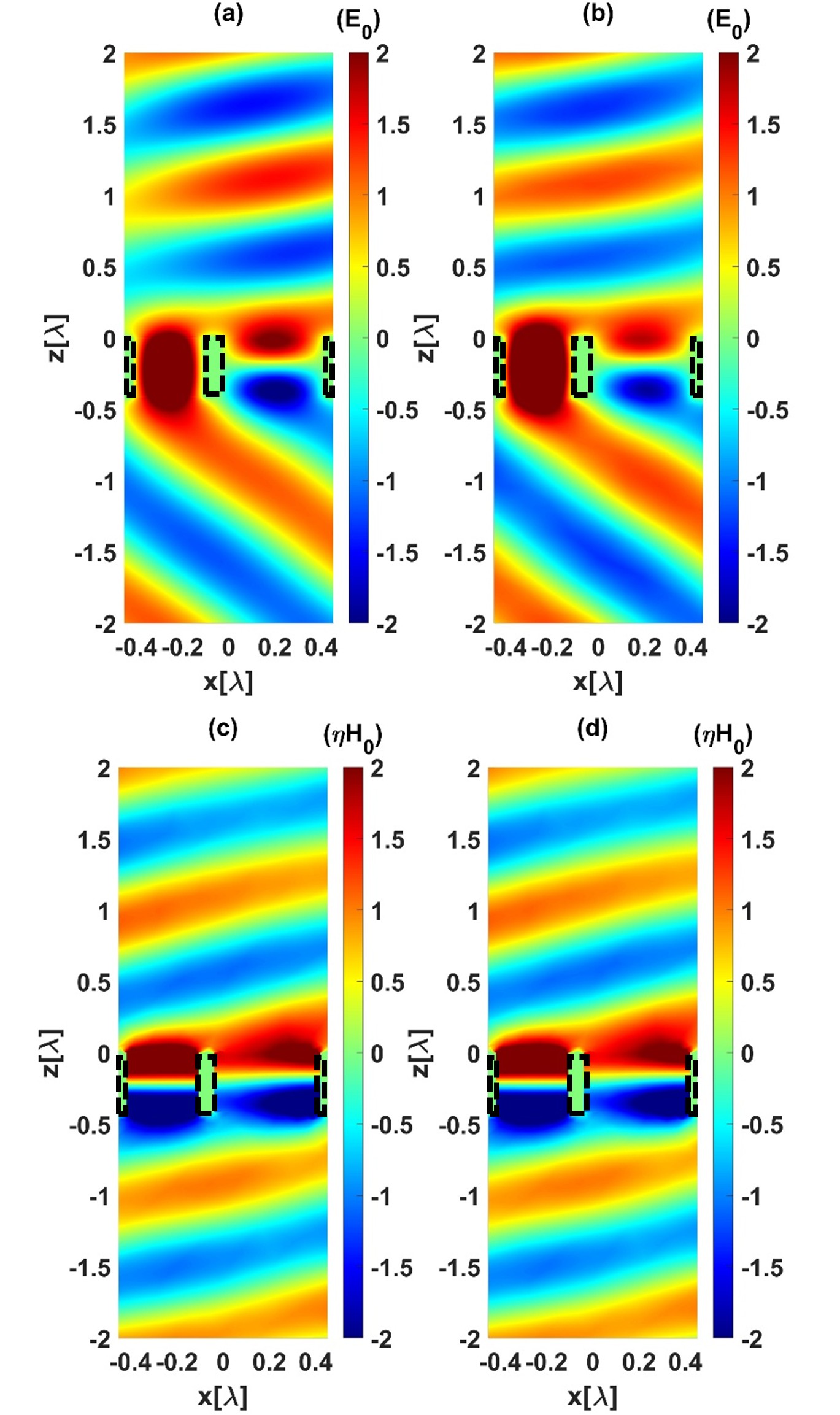}}
\caption{Electric field distributions for the dielectric-loaded polarization beam splitter (Section \ref{subsec:pbs}), when illuminated from $\theta_{\text{in}}=20\degree$ at $f=20\,\mathrm{GHz}$ (single period shown). $\text{Re}\{E_y(x,y=0,z)\}$ under TE-polarized incidence [(a),(b)] and $\text{Re}\{E_x(x,y=0,z)\}$ under TM-polarized incidence [(c),(d)]. Analytical predictions [(a),(c)] are compared with full-wave simulation results [(b),(d)]. Black dashed lines mark the slab boundaries.}
\label{fig:design3_fields}
\end{figure}

Complementing the efficiency assessment, Fig. \ref{fig:design3_fields} presents the spatial field distributions for both polarizations in the $\widehat{xz}$ plane at the design frequency. For TE-polarized incidence [Figs. \ref{fig:design3_fields}(a),(b)], $\text{Re}\{E_y(x,y=0,z)\}$ clearly shows the anomalously refracted transmitted wave at $\theta_{\text{out}} = -50\degree$ in the lower region ($z<-h$), distinct from the incident wave direction in the upper region ($z>0$). For TM-polarized excitation [Figs. \ref{fig:design3_fields}(c),(d)], the field snapshots reveal propagation through the MG into the fundamental ($(0,0)$-order) mode, with wavefronts in the lower region maintaining the same trajectory as the incident wave at $\theta_{\text{in}} = 20\degree$. The close agreement between analytical [(a),(c)] and full-wave [(b),(d)] results confirms that the mode-matching framework accurately captures the polarization-dependent routing in the devised MG configuration, resulting in an effective polarization beam splitter.

In fact, the results in Fig. \ref{fig:design3_fields} offer insights regarding the MG design to which the synthesis procedure converged [Fig. \ref{fig:TE1TM0_fabricated}(b)]. As discussed in the beginning of this subsection, since the considered MG utilizes dielectric loading, the reduced cutoff frequencies of the rectangular waveguides formed by the loaded perforations allows effective coupling of both TE- and TM-polarized incident waves to propagating waveguide modes. Examining the dominant field components of each polarization, we may conclude that TE-polarized fields (for which $E_x=0$) would predominantly couple to the $\mathrm{TE}_{m_x,m_y}=\mathrm{TE}_{1,0}$ waveguide mode, while TM-polarized incident fields ($E_y=0$) would mainly couple to the $\mathrm{TE}_{0,1}$ waveguide mode. Correspondingly, to cause fields emanating from TE-polarized excitation to accumulate different phases while passing through the $i=1$ and $i=2$ perforations -- required in order to modulate their phase fronts as to promote beam deflection [Fig. \ref{fig:design3_fields}(a), (b)] -- we need to change the $\mathrm{TE}_{1,0}$ propagation constant of the corresponding rectangular waveguides. As the latter are \emph{exclusively} controlled by the waveguide dimension along the $x$ axis, it follows that $d_x^{(1)}$ should be substantially different from $d_x^{(2)}$ to ensure efficient anomalous refraction. On the other hand, since TM-polarized wavefronts should propagate through the MG unaltered [Fig. \ref{fig:design3_fields}(c), (d)], it would make sense to devise the perforations such that they experience \emph{similar} phase gain upon propagation in the waveguides. Thus, one would expect similar longitudinal wavenumbers for the $\mathrm{TE}_{0,1}$ waveguide modes, leading to $d_y^{(1)}\approx d_y^{(2)}$ in the ultimate design. Indeed, reviewing the detailed MG parameters output by our synthesis scheme for this polarization-sensitive diffraction reveals quite distinct perforation dimensions along $x$, $d_x^{(1)}=4.64\,\mathrm{mm}\neq6.9\,\mathrm{mm}=d_x^{(2)}$, alongside closely matching cross sections along $y$, $d_y^{(1)}=7.0\,\mathrm{mm}\approx7.3\,\mathrm{mm}=d_y^{(2)}$ -- in consistency with the presented physical intuition. 

To validate experimentally the proposed concept and showcase its practical viability, a prototype was fabricated in the mechanical workshop at the Faculty of Electrical and Computer Engineering at the Technion, based on the layout devised and analyzed in Figs. \ref{fig:TE1TM0_fabricated}-\ref{fig:design3_fields}. This was carried out by CNC milling of a solid aluminum block, with the rectangular perforations machined to the specified dimensions. The Ultem 1000 dielectric inserts were fabricated as a periodically protruded structure -- the negative of the metallic frame's perforation pattern -- machined by CNC from a slightly thicker Ultem slab than the final design thickness $h$, leaving a thin planar Ultem layer connecting the protrusions. This insert was then bonded into the perforations of the machined aluminum slab using a standard epoxy adhesive (relative permittivity $\varepsilon_{r}\approx 3.2$, with a nominal bond-line thickness of $50\,\mu$m), and the resulting composite structure was given a final machining ("polishing") pass on both faces to remove residual material and ensure smooth, planar outer surfaces. The resulting assembly is shown in Fig. \ref{fig:TE1TM0_fabricated}(b).

\begin{figure}[!t]
\centerline{\includegraphics[width=\columnwidth]{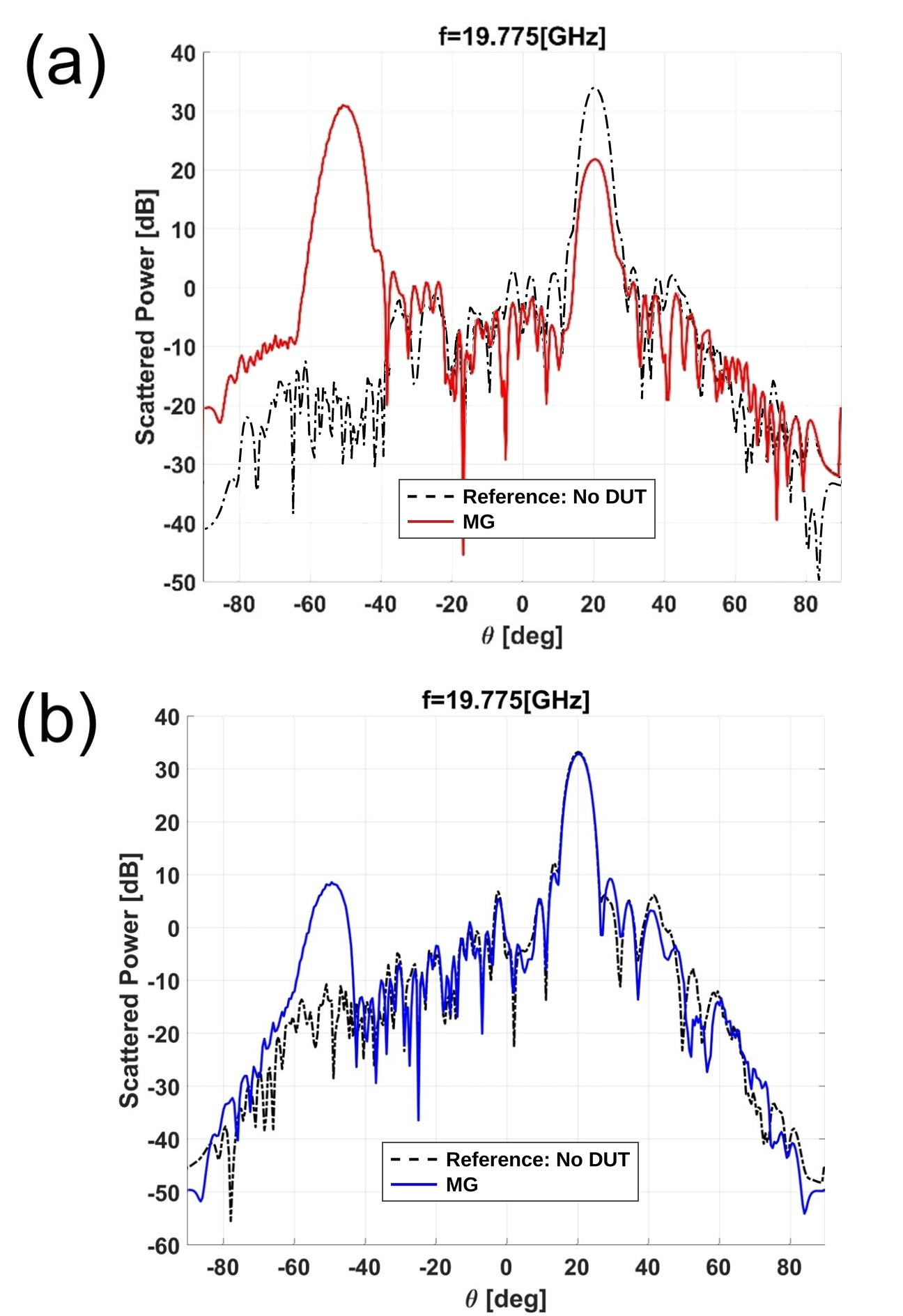}}
\caption{Measured far-field scattering patterns (solid lines) of the dielectric-loaded polarization beam splitter (Section \ref{subsec:pbs}) at the optimal operating frequency $f=19.775\,\mathrm{GHz}$, as a function of observation angle $\theta$, for (a) TE-polarized (red) and (b) TM-polarized (blue) excitation. The reference patterns recorded in the absence of the MG are shown as dashed lines.}
\label{fig:design3_patterns}
\end{figure}

\begin{figure}[!t]
\centerline{\includegraphics[width=1.0\columnwidth]{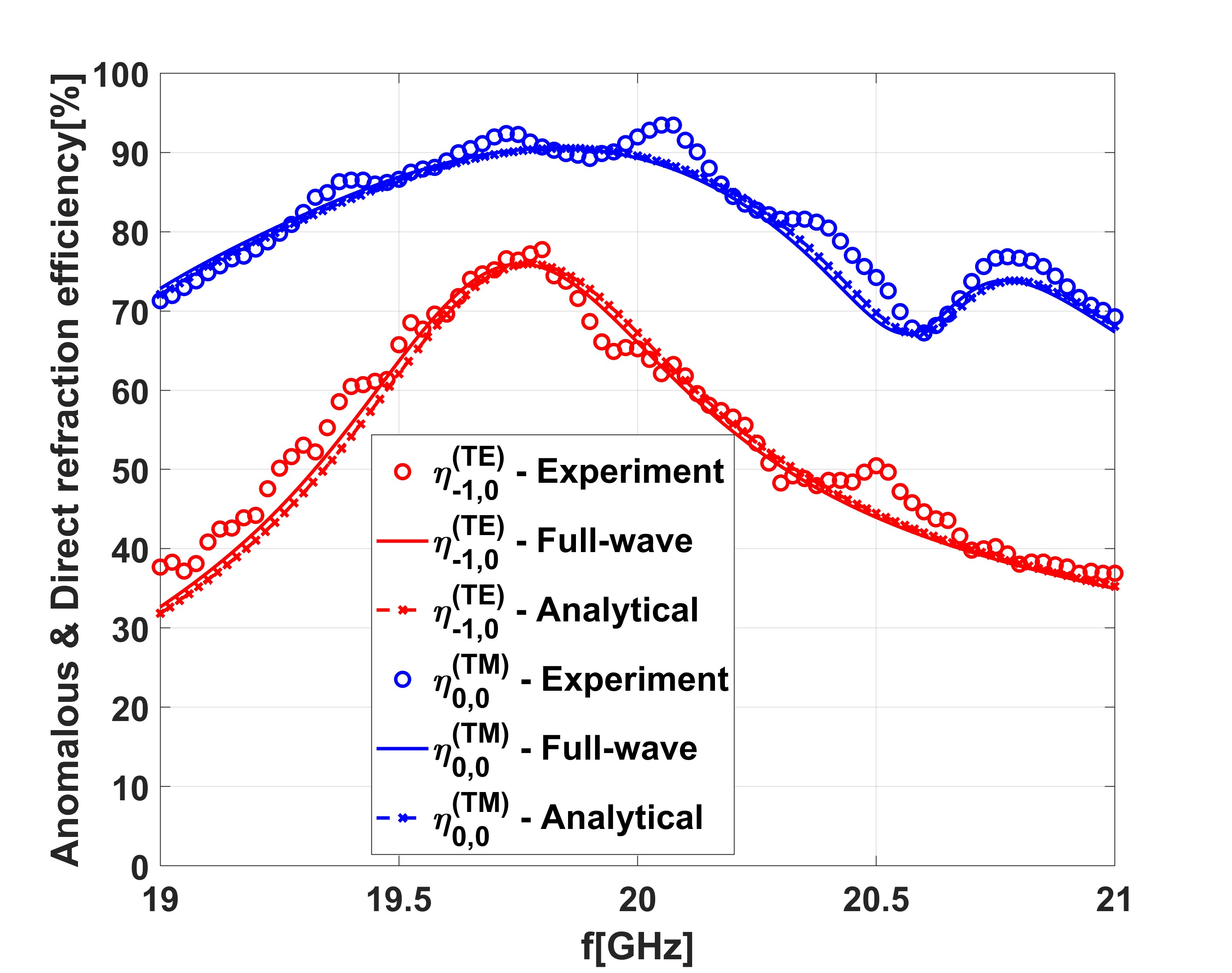}}
\caption{Measured anomalous refraction efficiencies of the dielectric-loaded polarization beam splitter (Section \ref{subsec:pbs}) as a function of frequency, quantifying the relative co-polarized power coupled to the $(-1,0)$ mode for TE-polarized incident fields (red) and to the fundamental $(0,0)$ mode for TM-polarized excitations (blue), as designated. Experimental results (circles) are compared with analytical predictions (dashed lines) and full-wave simulation results (solid lines), both evaluated at $\varepsilon_{d,r}=3.05$.}
\label{fig:design3_efficiency_TETM}
\end{figure}

It should be noted that due to the finite tool radius inherent to CNC machining, all rectangular perforation corners exhibit rounding with a radius of curvature of approximately $0.5$~mm, rather than the ideal sharp corners assumed in the analytical model. To preserve the target anomalous refraction efficiency in view of this constraint of the fabrication procedure, we redefined the MG configuration in CST Microwave Studio with this corner rounding explicitly incorporated, and made compensating adjustments to the perforation dimensions relative to the nominal design. Through a short optimization in CST, we have settled on the ultimate geometry for fabrication, restoring the goal functionality at the desired design frequency. 
For completeness, we present in Appendix \ref{appendix:fabrication} the design parameter values before and after these minor adjustments (Table \ref{tab:AppC1}), indicating variations of less than $4\%$ across all parameters \footnote{The analytical predictions and full-wave simulations compared in Figs. \ref{fig:design3_analytical}, \ref{fig:design3_fields}, \ref{fig:design4_efficiency}, and \ref{fig:design4_fields} are evaluated using the nominal (ideal, sharp-corner) perforation dimensions obtained from the synthesis procedure, ensuring a consistent, directly comparable basis between the two approaches. For comparison with the experimental measurements presented in Figs. \ref{fig:design3_patterns}, \ref{fig:design3_efficiency_TETM}, \ref{fig:design4_patterns}, and \ref{fig:design4_measured_efficiency}, a separate full-wave simulation was performed on the as-manufactured geometry that includes the rounded corners and the ultimate design parameters (Appendix \ref{appendix:fabrication}).}. 

\begin{figure}[!t]
\centerline{\includegraphics[width=0.8\columnwidth]{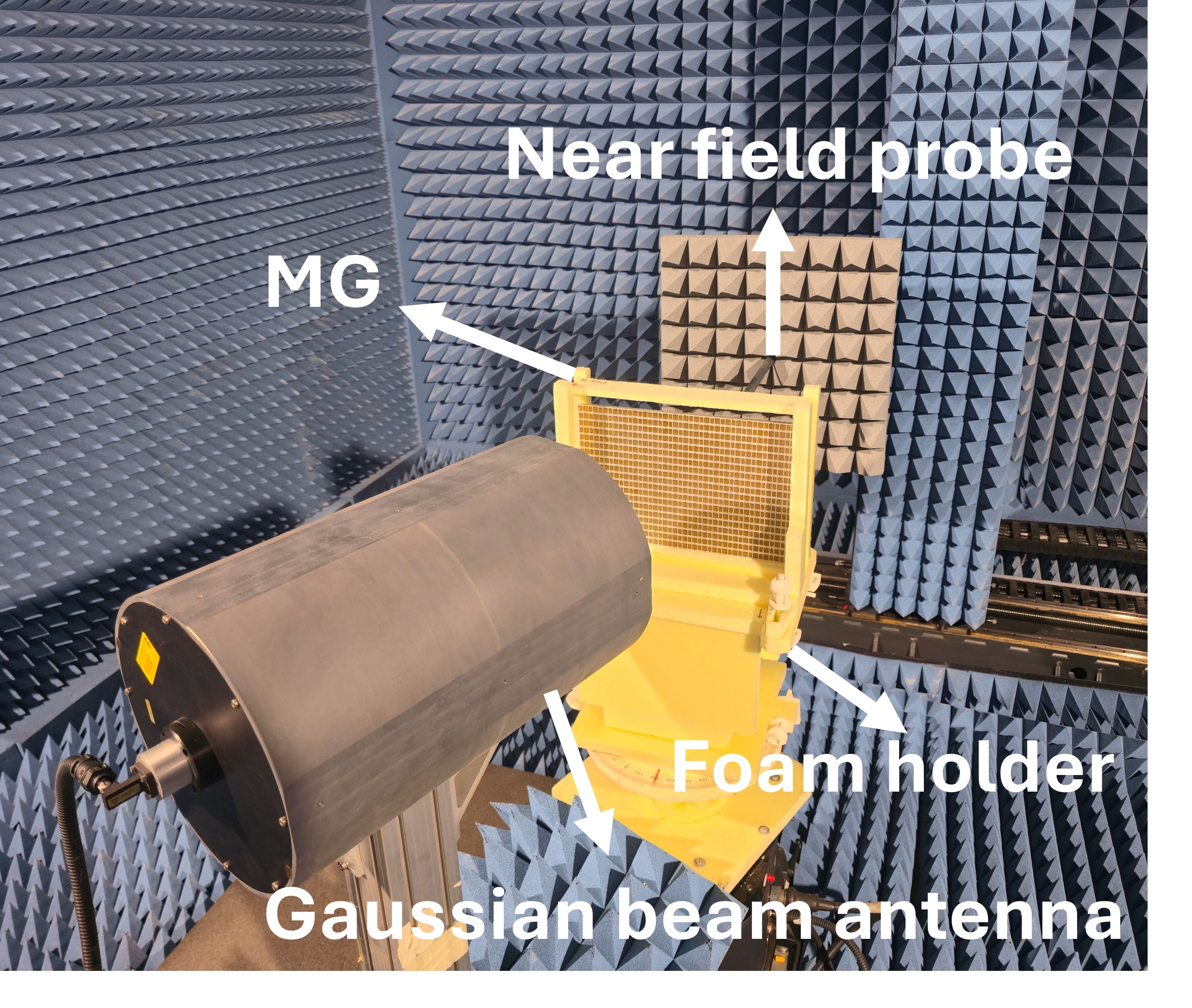}}
\caption{Anechoic chamber measurement setup used for experimental characterization of the dielectric-loaded MG prototypes (Sections \ref{subsec:pbs} and \ref{subsec:dual}). The fabricated MG prototype is mounted on a foam holder and positioned between a Gaussian beam antenna (transmitting source) and a near-field scanning probe.}
\label{fig:chamber}
\end{figure}

Subsequently, we moved the fabricated prototype to the anechoic chamber at the Technion for characterizing its scattering performance via the installed near-field measurement system (MVG/Orbit-FR). Within the measurement setup (Fig. \ref{fig:chamber}), a Gaussian beam antenna (Millitech, Inc., GOA-42-S000094) was positioned at a distance of $196$~mm $\approx 13\lambda$ from the MG (mounted on a foam holder), aligned to illuminate it by quasi-plane wave excitation at an angle of incidence $\theta_{\text{in}}=20^\circ$. An open-ended waveguide (OEW) probe was used to scan the plane $z=245$~mm ($\approx 16.3\lambda$) from the MG within the range $x\in[-500,500]$~mm, $y\in[-200,200]$~mm, recording the amplitude and phase of the transmitted electric fields. Finally, application of the equivalence principle \cite{Balanis2012} allows calculation of the far-field scattering pattern, from which the fraction of incident power coupled to the various transmitted FB harmonics (coupling efficiencies) can be deduced. 

In particular, the anomalous refraction efficiency is evaluated from the recorded frequency-resolved gain $G(\theta;f)$ as in \cite{Wong2018,Rabinovich2020,DiazRubio2017}, following
\begin{equation}
\eta^{\{\mathrm{TE,TM}\}}_{-1,0}(f) = \frac{G^{\{\mathrm{TE,TM}\}}_{\text{MG}}(\theta_{\text{out}};f)}{G^{\{\mathrm{TE,TM}\}}_{\text{direct}}(\theta_{\text{in}};f)}\cdot\frac{\cos\theta_{\text{in}}}{\cos\theta_{\text{out}}},
\label{eq:efficiency_from_gain}
\end{equation}
where $G_{\text{MG}}(\theta_{\text{out}};f)$ is the peak gain measured by the OEW probe with the MG in place at the frequency $f$ at the anomalous refraction angle $\theta_{\text{out}}$, and $G_{\text{direct}}(\theta_{\text{in}};f)$ is the reference gain measured at the angle of incidence $\theta_{\mathrm{in}}$ and the same frequency in the absence of the MG. Since both the exciting Gaussian beam antenna and the OEW probe are single-polarized, the TE- and TM- polarized scattering patterns are characterized sequentially, with the transmitting antenna and probe both rotated by $90\degree$ between the two measurement sets.

The measured far-field scattering patterns for the prototype of Fig. \ref{fig:TE1TM0_fabricated} are presented in Fig. \ref{fig:design3_patterns}, for the optimal operating frequency found in the experiment ($f=19.775\,\mathrm{GHz}$).
Indeed, the results provide a qualitative validation of the orthogonal transmission functionality: the scattering due to TE-polarized excitation [Fig. \ref{fig:design3_patterns}(a), red solid lines] shows a clear peak at $\theta_{\text{out}} = -50\degree$ and suppressed direct transmission (around $+20\degree$) with respect to the reference (dashed black lines); while the co-polarized scattered power obtained for TM-polarized incident fields [Fig. \ref{fig:design3_patterns}(b), solid blue] maintains almost unperturbed transmission ($\theta_{\mathrm{in}}=20\degree$) with significantly reduced power at the spurious anomalous refraction direction, confirming polarization-dependent beam routing.

To quantify the performance, Fig. \ref{fig:design3_efficiency_TETM} presents the experimentally evaluated coupling efficiencies to the relevant transmitted FB modes as a function of frequency, corresponding to the dominant diffraction orders for each of the considered polarizations. For TE-polarized field excitation, the measured fraction of TE-polarized power coupled to the $(-1,0)$-order mode (red circle markers) reveals a peak efficiency of $\eta_{-1,0}^{\text{TE}}=77\%$ at $19.775\,\mathrm{GHz}$, reflecting a downward frequency shift of approximately $1.0\%$ relative to the nominal design frequency (already noted in the previous paragraph). For the TM polarization, the recorded direct transmission coupling efficiency $\eta_{0,0}^{\text{TM}}$ (blue circle markers) experiences the same frequency shift, peaking at a higher value of about $91\%$.

While the prototype performs reasonably well as the transmissive polarization beam splitter envisioned near the designated operating frequency, the somewhat reduced coupling efficiencies and mild frequency shift compared to the original predictions (\textit{cf.} Fig. \ref{fig:design3_analytical}) require a dedicated discussion. Considering the typical dielectric constant variation expected for extruded/injection-molded polymeric dielectrics of this kind (commonly on the order of a few percent around the nominal value), a downward frequency shift may point out a slight increase in the relative permittivity in the realized prototype with respect to the nominal design value. To probe this hypothesis, we repeated our calculations with the analytical model and reran the full-wave simulations for the same MG configuration, but now with the loading permittivity taken to a slightly increased value, $\varepsilon_{d,r} = 3.05$ \footnote{As denoted in \cite{Note3}, the simulated configuration for the comparison with the experimental results in Fig. \ref{fig:design3_efficiency_TETM} already considers the corner rounding and minor parameter adjustments previously obtained, as documented in Table \ref{tab:AppC1}.}. The resultant coupling efficiencies, presented in Fig. \ref{fig:design3_efficiency_TETM} in dash-dotted (analytical) and solid (full-wave) lines for the TE- (red) and TM- (blue) polarized cases exhibit remarkable correspondence with the experimental data (both in terms of frequency shift and coupling coefficients), thereby confirming that the dielectric constant variation is indeed the primary source of the observed discrepancy.

Overall, these results validate the capability of dielectric-loaded metallic MGs to implement effective polarization-dependent beam routing within a compact single-layer configuration in practice, overcoming the limitations associated with TE-polarized field manipulation discussed in Section \ref{subsec:te_double} while providing insights regarding the solution geometry.

\subsection{Single-Layer Dielectric-Loaded MG for Dual-Polarized Anomalous Refraction}
\label{subsec:dual}

To further exemplify the versatility of our approach, and, specifically, the diverse dual-polarized control functionalities realizable with the single-layer dielectric-loaded metallic MGs introduced in Section \ref{subsec:pbs}, we aim in this section to devise and characterize such a composite for polarization-\emph{insensitive} diffraction engineering. In particular, in contrast to separating the polarization components as implemented by the MG presented in the previous section (essentially, without interfering with TM-polarized field propagation), our synthesis goal herein would be to achieve simultaneous dual-polarized anomalous refraction to the same $(-1,0)$ diffraction order (same angle), requiring trajectory-manipulating interactions with both TE- and TM-polarized excitations. Such MGs may, in principle, be useful for applications addressing circularly polarized signals \cite{Tong2023,Yang2024}, or systems where electromagnetic power needs to be funnelled regardless of its polarization properties, e.g., dual-polarized RF energy harvesting devices \cite{Sun2020}.

\begin{figure}[!t]
\centerline{\includegraphics[width=1.0\columnwidth]{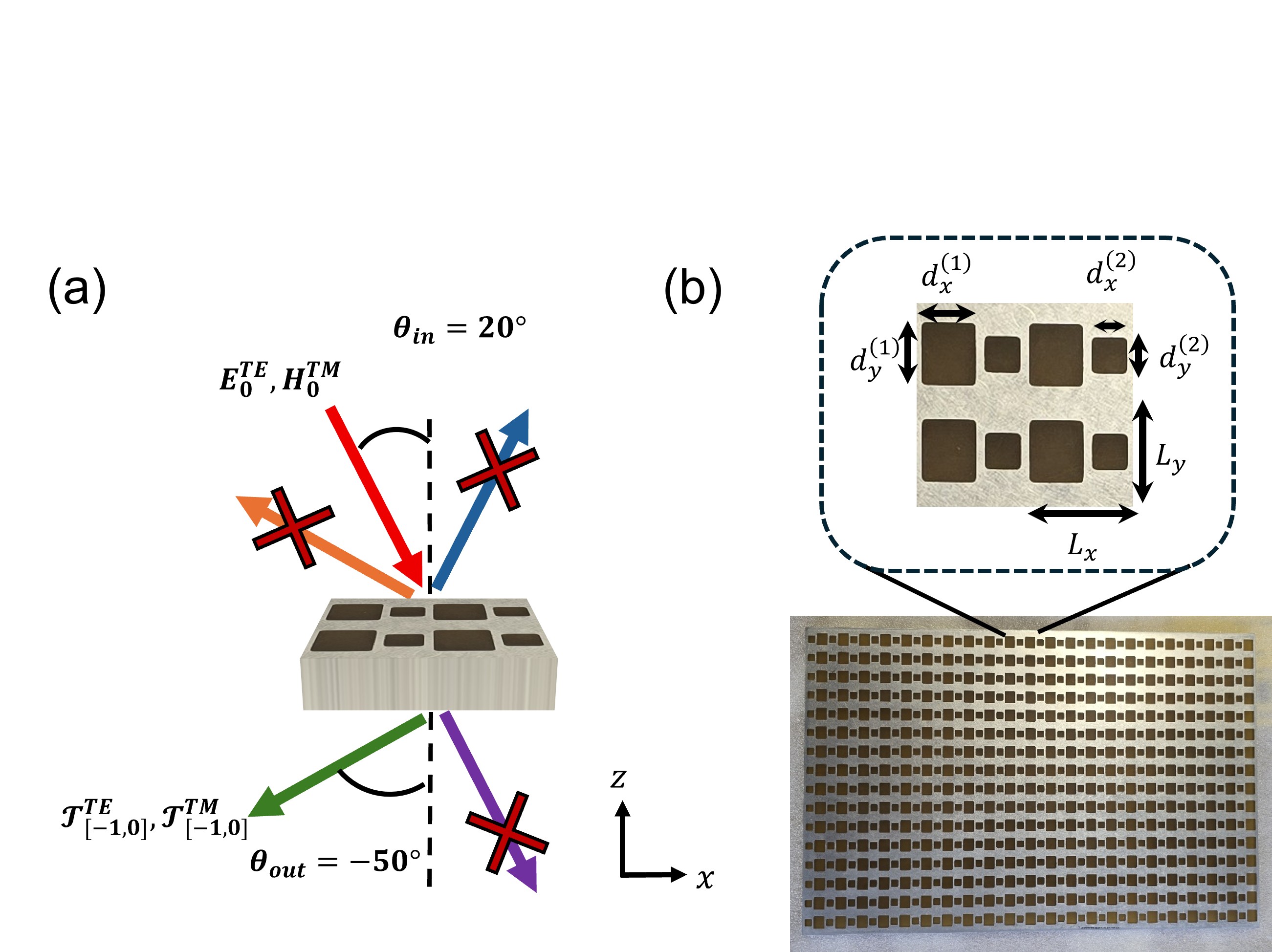}}
\caption{Dielectric-loaded metallic-based MG for co-directed dual-polarized anomalous refraction (Section \ref{subsec:dual}). (a) Three-dimensional scattering scenario, illustrating the target dual-polarized anomalous refraction functionality: both TE- and TM-polarized incident fields from $\theta_{\mathrm{in}}=20^\circ$ are redirected towards $\theta_{\text{out}}=-50^\circ$. (b) Full view of the fabricated prototype: CNC-machined aluminum panel with Ultem 1000 dielectric inserts; inset: closeup on the unit cell, with the relevant geometric parameters indicated.}
\label{fig:TE1TM1_fabricated}
\end{figure}

The target functionality is therefore presented in Fig. \ref{fig:TE1TM1_fabricated}(a), where the corresponding synthesis goal invoked as part of the procedure outlined in Section \ref{subsec:synthesis} involves maximizing the anomalous refraction efficiencies for both polarizations simultaneously, namely, $\eta_{-1,0}^{\text{TE}} \to 1$ and $\eta_{-1,0}^{\text{TM}} \to 1$. 
Following the same rationale as in Section \ref{subsec:pbs}, we choose to approach the design task using $N_p = 2$ dielectric-loaded perforations per MG period, which are expected to provide the required degrees of freedom for the intended function. Applying once more the same inverse solution scheme (Section \ref{subsec:synthesis}) with the analytical model of Section \ref{subsec:single_layer} yields the following optimal parameters for the slab thickness, perforation geometry, and MG periodicity along $y$: $h = 5.8\,\mathrm{mm}$; $d_x^{(1)} = 4.6\,\mathrm{mm}$, $d_y^{(1)} = 4.7\,\mathrm{mm}$, and $(a_x^{(1)}, a_y^{(1)}) = (-3.98\,\mathrm{mm}, 0)$; $d_x^{(2)} = 7.0\,\mathrm{mm}$, $d_y^{(2)} = 7.9\,\mathrm{mm}$, and $(a_x^{(2)}, a_y^{(2)}) = (2.79\,\mathrm{mm}, 0)$; and $L_y = 13.2\,\mathrm{mm}$ [Fig. \ref{fig:TE1TM1_fabricated}(b)].

\begin{figure}[!t]
\centerline{\includegraphics[width=1.0\columnwidth]{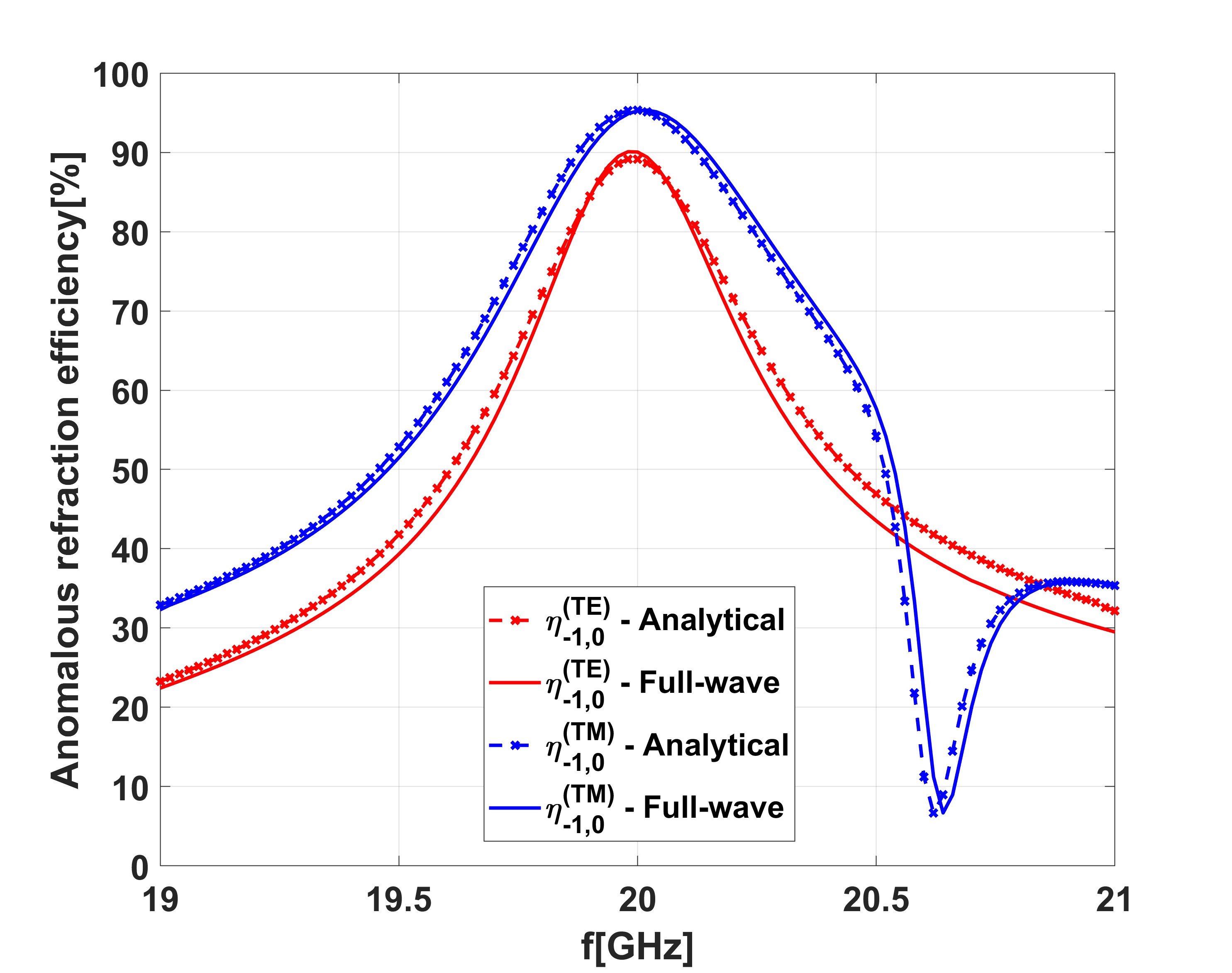}}
\caption{Anomalous refraction efficiencies of the dual-polarized dielectric-loaded MG (Section \ref{subsec:dual}) as a function of frequency, quantifying the relative co-polarized power coupled to the $(-1,0)$ mode for TE- (red) and TM- (blue) polarized incident fields, as designated. Analytical predictions (dashed lines) are compared with full-wave simulation results (solid lines).}
\label{fig:design4_efficiency}
\end{figure}

The resulting dual-polarized performance is captured in Fig. \ref{fig:design4_efficiency}, which overlays the analytically predicted (dashed lines with markers) and full-wave simulated (solid lines) $\eta_{-1,0}^{\text{TE}}$ and $\eta_{-1,0}^{\text{TM}}$ across frequency for the devised MG layout. Excellent agreement is observed, with both sets of curves peaking at the design frequency of $20\,\mathrm{GHz}$, indicating that $97\%$ of the TM-polarized incident power (blue) and $90\%$ of the impinging TE-polarized radiation (red) are effectively redirected to the anomalous refraction mode.

\begin{figure}[!t]
\centerline{\includegraphics[width=0.8\columnwidth]{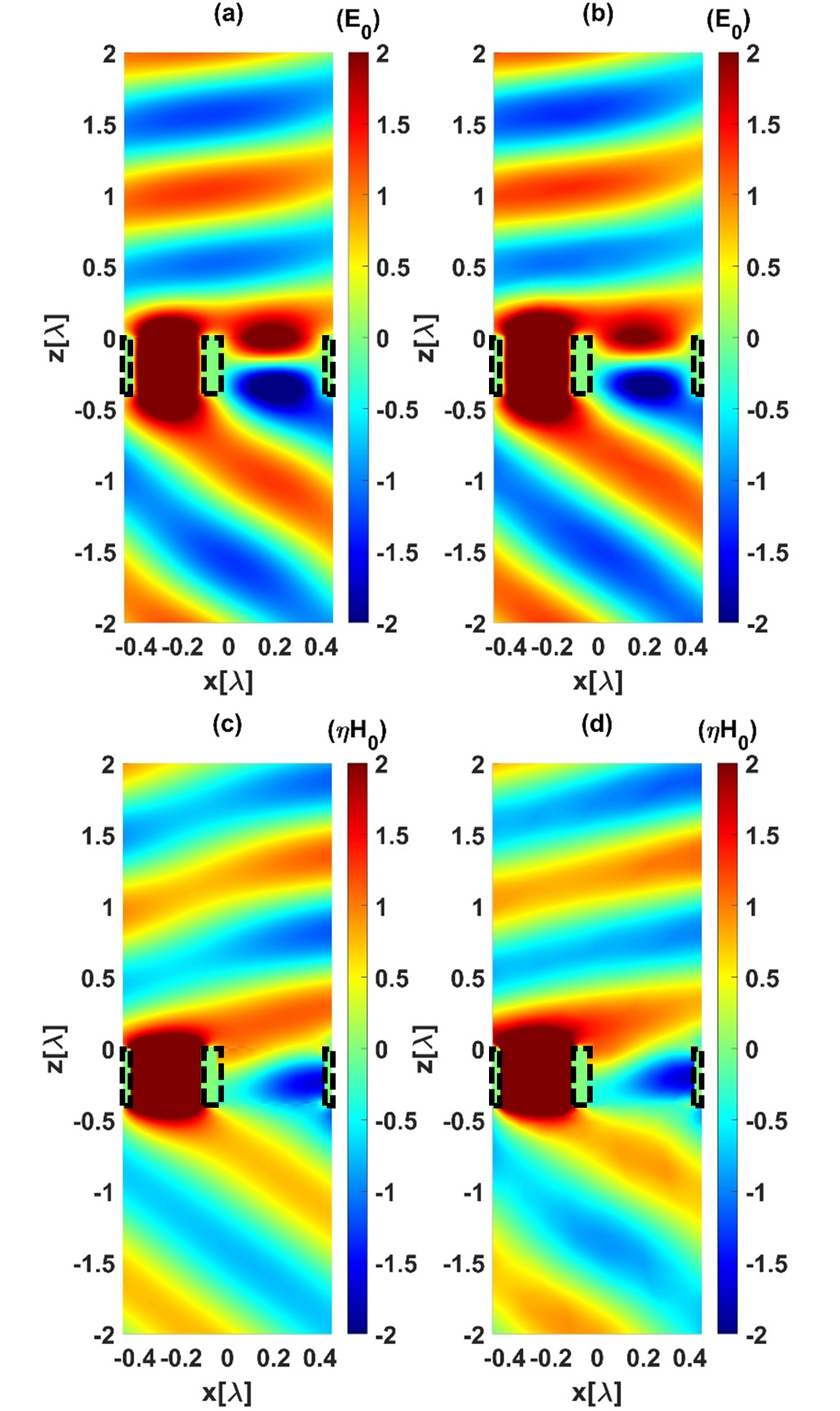}}
\caption{Electric field distributions for the dual-polarized dielectric-loaded MG (Section \ref{subsec:dual}), when illuminated from $\theta_{\text{in}}=20\degree$ at $f=20\,\mathrm{GHz}$ (single period shown). $\text{Re}\{E_y(x,y=0,z)\}$ under TE-polarized incidence [(a),(b)] and $\text{Re}\{E_x(x,y=0,z)\}$ under TM-polarized incidence [(c),(d)]. Analytical predictions [(a),(c)] are compared with full-wave simulation results [(b),(d)]. Black dashed lines mark the slab boundaries.}
\label{fig:design4_fields}
\end{figure}

The spatial field distributions in the $\widehat{xz}$ plane, shown in Fig. \ref{fig:design4_fields} for both polarizations at $20\,\mathrm{GHz}$, provide additional visual confirmation of this dual-polarized operation. For both TE- [Figs. \ref{fig:design4_fields}(a),(b)] and TM- [Figs. \ref{fig:design4_fields}(c),(d)] polarized incidence from $\theta_\mathrm{in}=20^\circ$ ($z>0$), the transmitted wave is clearly being steered toward $\theta_{\text{out}} = -50\degree$ in the lower region ($z<-h$), as desired. The close agreement between analytical [Figs. \ref{fig:design4_fields}(a),(c)] and full-wave [Figs. \ref{fig:design4_fields}(b),(d)] results validates the framework for this functionality. In addition, examining these plots in view of the ultimate design of Fig. \ref{fig:TE1TM1_fabricated}(b) allows us to reinforce the intuitive observations discussed in relation to Fig. \ref{fig:design3_fields} in Section \ref{subsec:pbs}. When implementing anomalous refraction for both polarization components, both TE- and TM- polarized fields in the incident wave should experience a spatially varying phase shift upon transmission through the MG. In other words, the propagation constants of the dominantly excited rectangular waveguide modes in each of the perforations should differ substantially -- for either excitation. Indeed, the optimal perforation cross sections dictated by the synthesis algorithm and listed in this subsection -- with $d_x^{(1)}\neq d_x^{(2)}$ and $d_y^{(1)}\neq d_y^{(2)}$ [Fig. \ref{fig:TE1TM1_fabricated}(b)] -- clearly lead to significantly different phase accumulation in perforation $i=1$ and perforation $i=2$ for both TE-polarized incident components (predominantly coupled to the propagating $\mathrm{TE}_{0,1}$ waveguide modes) as well as the TM-polarized ones ($\mathrm{TE}_{1,0}$ waveguide modes). This is in contrast to the design in Fig. \ref{fig:TE1TM0_fabricated}(b), where $d_y^{(1)}\approx d_y^{(2)}$ due to the requirement for (uninterrupted) direct transmission of TM-polarized excitations applicable to the MG designed in Section \ref{subsec:pbs}, necessary to establish the prescribed polarization selectivity (see also the relevant discussion therein). Interestingly, the distinct profiles of the $\mathrm{TE}_{0,1}$ and $\mathrm{TE}_{1,0}$ waveguide modes in the $\widehat{xz}$ plane can be clearly observed in the $i=2$ perforations of Figs. \ref{fig:design4_fields}(a),(b) and Figs. \ref{fig:design4_fields}(c),(d), respectively, further supporting our conclusions.

Encouraged by these theoretical results, we proceeded towards fabrication of a corresponding prototype, realizing the configuration corresponding to Figs. \ref{fig:design4_efficiency} and \ref{fig:design4_fields} following the same CNC milling and dielectric insert procedure described in Section \ref{subsec:pbs}. As before, the finite CNC tool radius introduces rounding of approximately $0.5$~mm at all rectangular perforation corners, requiring minor compensation to the perforation dimensions with respect to the original design. The corresponding MG parameters before and after these adjustments are listed in Table \ref{tab:AppC2} of Appendix \ref{appendix:fabrication}. 
The finalized composite, shown in Fig. \ref{fig:TE1TM1_fabricated}(b), was subsequently characterized using the setup of Fig. \ref{fig:chamber}, with the dual-polarized response evaluated from separate single-polarized measurements, as described in Section \ref{subsec:pbs} as well \cite{Note3}.

\begin{figure}[!t]
\centerline{\includegraphics[width=\columnwidth]{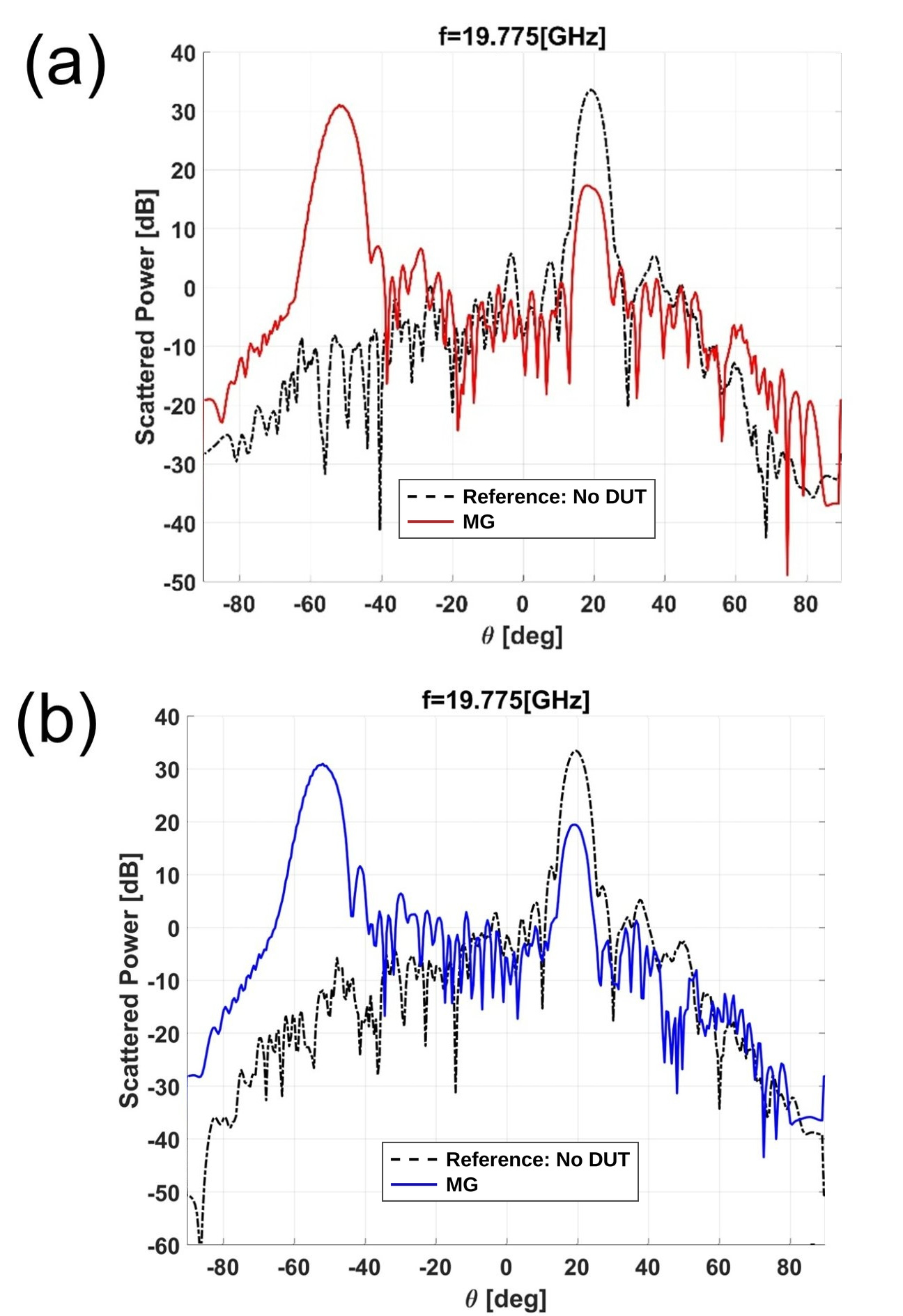}}
\caption{Measured far-field scattering patterns of the dual-polarized dielectric-loaded MG (Section \ref{subsec:dual}) at the optimal operating frequency $f=19.775\,\mathrm{GHz}$, as a function of observation angle $\theta$, for (a) TE-polarized (red) and (b) TM-polarized (blue) excitation. The reference patterns recorded in the absence of the MG are shown as dashed lines.}
\label{fig:design4_patterns}
\end{figure}

The measured scattering patterns presented in Fig. \ref{fig:design4_patterns} for the optimal operating frequency found experimentally ($f=19.775\,\mathrm{GHz}$, as in Section \ref{subsec:pbs}) confirm that, indeed, both TE- [Fig. \ref{fig:design4_patterns}(a)] and TM-[Fig. \ref{fig:design4_patterns}(b)] polarized excitations of the manufactured MG prototype lead to co-polarized anomalous refraction towards the common $\theta_{\text{out}}\approx-50^\circ$, as designed.

\begin{figure}[!t]
\centerline{\includegraphics[width=1.0\columnwidth]{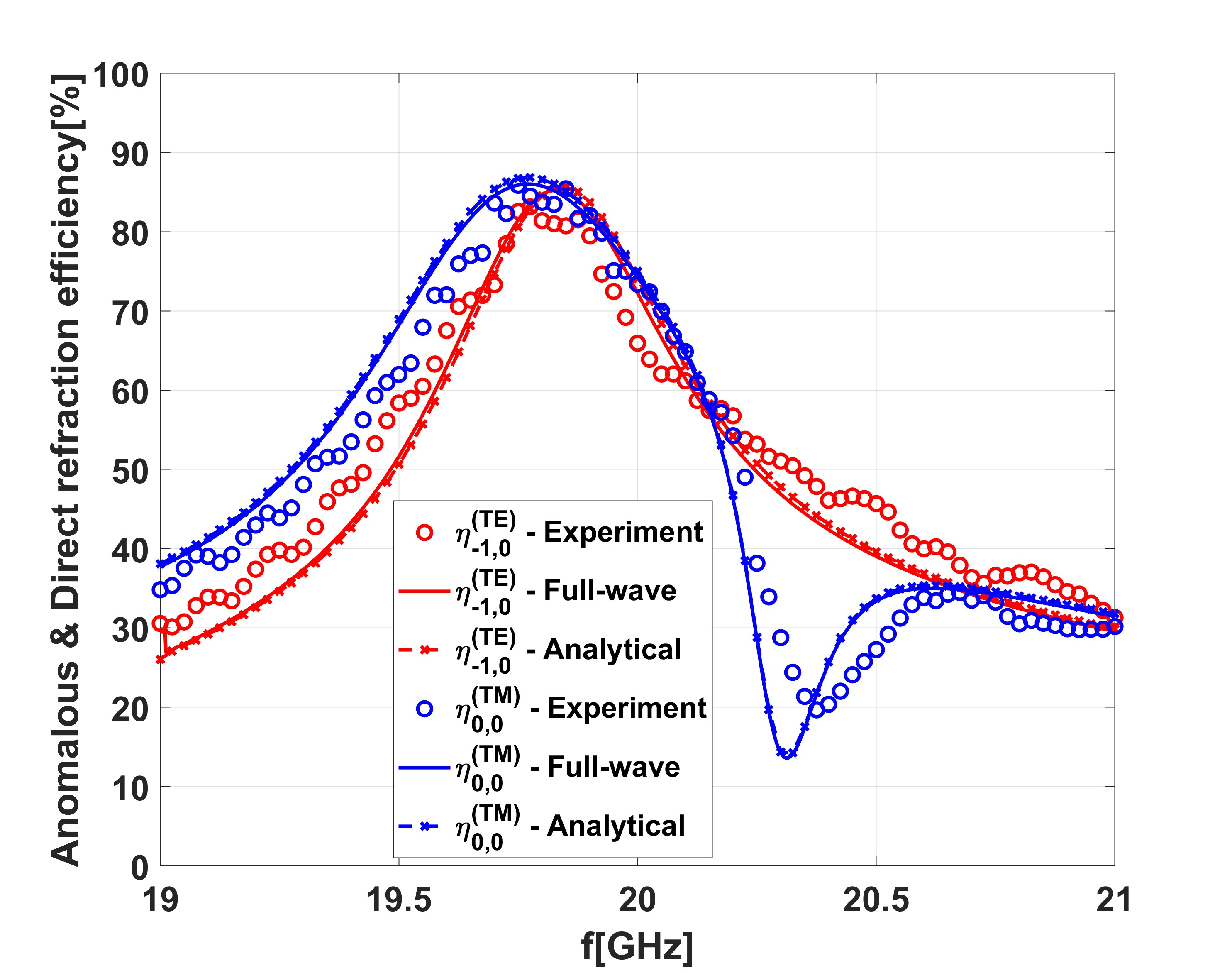}}
\caption{Measured anomalous refraction efficiencies of the dual-polarized dielectric-loaded MG (Section \ref{subsec:dual}) as a function of frequency, quantifying the relative co-polarized power coupled to the $(-1,0)$ mode for TE- (red) and TM- (blue) polarized incident fields, as designated. Experimental results (circles) are compared with analytical predictions (dashed lines) and full-wave simulation results (solid lines), both evaluated at $\varepsilon_{d,r}=3.05$.}
\label{fig:design4_measured_efficiency}
\end{figure}

Beyond this qualitative picture, the experimentally evaluated coupling efficiencies into the $(-1,0)$-order FB mode in transmission (circle markers), presented in Fig. \ref{fig:design4_measured_efficiency} for both incident field polarization components, allow for a quantitative comparison with the theoretical predictions of Fig. \ref{fig:design4_efficiency}. We again identify the same $\sim1\%$ frequency shift and $\sim10\%$ efficiency drop as seen in the experimental results obtained for the polarization beam splitting MG (Fig. \ref{fig:design3_efficiency_TETM}), with anomalous refraction efficiencies peaking at $19.775\,\mathrm{GHz}$ with values of $87\%$ and $85\%$ for TM- (blue) and TE- (red) polarized incident fields, respectively. Equipped with our observations from Section \ref{subsec:pbs}, we attribute this discrepancy to the same slight upward deviation in the realized Ultem 1000 permittivity discussed therein. Repeating the analytical calculation (dash-dotted lines) and full-wave simulations (solid lines) with $\varepsilon_{d,r}=3.05$ in Fig. \ref{fig:design4_measured_efficiency}) again restores the agreement with the measured results, corroborating this explanation.

These theoretical and experimental demonstrations combined indicate that even in view of realistic fabrication limitations and material tolerances, the devised architectures succeed in implementing highly efficient polarization (in)sensitive beam deflection in transmission at application-relevant K band frequencies, showcasing the practical viability of dielectric-loaded metallic MGs for advanced dual-polarized beamforming operations.

\section{Conclusion}
\label{sec:conclusion}

To summarize, we introduced a rigorous mode-matching framework for the design of transmissive metallic metagratings, extending Floquet-Bloch analysis to perforated slabs -- single- or double-layer, with an arbitrary number of rectangular, optionally dielectric-loaded apertures per period -- providing direct analytical access to the coupling coefficients for both TE- and TM-polarized fields. Since the model ties the scattering coefficients explicitly to the perforation geometry, it supports a genuine inverse-design procedure, translating prescribed diffraction requirements directly into fabrication-ready MG layouts without resorting to full-wave optimization loops. To demonstrate the versatility of our approach, we devised four different metallic-based MG designs for beam deflection -- from all-metallic single-polarized single-layer and double-layer layouts to dielectric-loaded slabs for dual-polarized field manipulation -- showcasing through full-wave simulations and selective laboratory experiments the functional efficacy and practical viability of the proposed concept. 

Across the case studies, a recurring observation was the asymmetry between TM- and TE-polarized field manipulation in this particular metallic waveguide-based topology, rooted in the different cutoff behavior of the underlying rectangular-waveguide modes: while TM-polarized beam steering was readily achieved with a single perforated layer, effective TE-polarized control required one of two remedies. The first, a double-layer construct with an intermediate air gap, harnesses multiple reflections and evanescent-mode tunnelling to compensate for the lack of usable phase gradients within a single metallic layer, but at the cost of a narrower operational bandwidth and an increased risk of unwanted lateral power confinement. The second, dielectric loading of the perforations, directly lowers the relevant modal cutoff frequencies, restoring access to propagating modes for both polarizations simultaneously within a single compact layer. This latter route proved the more practical of the two, forming the basis for the two designs selected for experimental validation. 

Overall, the validated framework, together with the physical insights accompanying it, offers a systematic path toward metallic-based, mechanically robust and power-resilient composite MG superstrates for transmissive, dual-polarized beamforming in demanding environments -- of particular relevance for spaceborne, high-power, and high-frequency systems.

\appendix

\section{Boundary Condition Coefficients\\and Overlap Integrals}
\label{appendix:coefficients}
In this appendix, we provide explicit expressions for the various coefficients and overlap integrals introduced in Section \ref{subsec:single_layer}, required for the stipulation and resolution of the boundary conditions associated with the transmissive MG configuration considered herein. The coefficients $\alpha_{\mathbf{n}}^{(q)}$, $\beta_{\mathbf{m}}^{(q,i)}$, $\beta_{\mathbf{m}^*}^{(q,i)}$, and $S_{\mathbf{n}}^{(q)}$ (for $q=\text{I,II}$) or $S_{\mathbf{m}}^{(q,i)}$ (for $q=\text{III,IV}$) appearing in Eqs. \eqref{eq:Ex_upper}-\eqref{eq:Hy_lower} are obtained by substituting the longitudinal field components into Maxwell's equations to derive their tangential counterparts, and subsequently applying the mode-matching procedure detailed in Section \ref{subsec:single_layer}, following the approach of \cite{Rabinovich2020_1}.

For the electric field boundary conditions Eqs. \eqref{eq:Ex_upper}-\eqref{eq:Ey_lower}, multiplying by $e^{jk_x^{(n_x)}x}e^{jk_y^{(n_y)}y}$ and integrating over the unit cell $L_x\times L_y$ \cite{Rabinovich2020_1} yields the overlap integrals $\psi^{(i)}_{\mathbf{n},\mathbf{m}}$ and $\chi^{(i)}_{\mathbf{n},\mathbf{m}}$ between the $(n_x,n_y)$-order FB mode and the $(m_x,m_y)$-order waveguide mode, where the integration reduces to the perforation cross-sectional area since the waveguide fields vanish outside their respective apertures:
\begin{equation}
\begin{split}
\psi^{(i)}_{\mathbf{n},\mathbf{m}}&=\int_{y_-^{(i)}}^{y_+^{(i)}}\int_{x_-^{(i)}}^{x_+^{(i)}}\cos\left[\frac{\pi m_x}{d_x^{(i)}}(x - x_-^{(i)})\right]\\
&\quad\times\sin\left[\frac{\pi m_y}{d_y^{(i)}}(y - y_-^{(i)})\right]e^{jk_x^{(n_x)} x} e^{jk_y^{(n_y)} y}\, dxdy
\end{split}
\label{eq:psi_appendix}
\end{equation}
and
\begin{equation}
\begin{split}
\chi^{(i)}_{\mathbf{n},\mathbf{m}}&=\int_{y_-^{(i)}}^{y_+^{(i)}}\int_{x_-^{(i)}}^{x_+^{(i)}}\sin\left[\frac{\pi m_x}{d_x^{(i)}}(x - x_-^{(i)})\right]\\
&\quad\times\cos\left[\frac{\pi m_y}{d_y^{(i)}}(y - y_-^{(i)})\right]e^{jk_x^{(n_x)} x} e^{jk_y^{(n_y)} y}\, dxdy,
\end{split}
\label{eq:chi_appendix}
\end{equation}
both evaluated analytically using standard integral identities \cite{Rabinovich2020_1}. For the magnetic field boundary conditions Eqs. \eqref{eq:Hx_upper}-\eqref{eq:Hy_lower}, FB mode orthogonality cannot be exploited since continuity is enforced only over the perforation apertures. Instead, the waveguide eigenmode orthogonality is applied by multiplying by the appropriate eigenfunction and integrating over each perforation aperture separately \cite{Balanis2012}. The resulting modal coefficients $\alpha_{\mathbf{n}}^{\text{TM}/\text{TE}(q)}$, $\beta^{\text{TM}(q,i)}_{\mathbf{m}}$ and $\beta_{\mathbf{m}^*}^{\text{TE}(q,i)}$ are collected in Table \ref{tab:coefficients}, along with the source terms $S_{\mathbf{n}}^{(q)}$ (for $q=\text{I,II}$) and $S_{\mathbf{m}}^{(q,i)}$ (for $q=\text{III,IV}$), expressed in terms of the polarization-dependent source amplitudes $s^{(q)}$ as:
\begin{align}
\begin{aligned}
s^{(\text{I})} &= \eta H_0^{\text{TM}} \cos\theta_{\text{in}},\\ 
s^{(\text{II})} &= E_0^{\text{TE}}, \\
s^{(\text{III})} &= E_0^{\text{TE}} \cos\theta_{\text{in}}, \\ 
s^{(\text{IV})} &= -\eta H_0^{\text{TM}}.
\end{aligned}
\label{eq:source_terms}
\end{align}
Here, $s^{(\text{I})}$ and $s^{(\text{IV})}$ are the $E_x$- and $H_y$-component source amplitudes of the TM-polarized incident field ($H_0^{\text{TM}}$), respectively, while $s^{(\text{II})}$ and $s^{(\text{III})}$ are the $E_y$- and $H_x$-component source amplitudes of the TE-polarized incident field ($E_0^{\text{TE}}$), respectively; the remaining counterparts vanish identically for each polarization. All source amplitudes are evaluated for the incident fundamental $(0,0)$-order FB mode.

Once the coefficients in Eqs. \eqref{eq:Ex_upper}-\eqref{eq:Hy_lower} are established through Eqs. \eqref{eq:psi_appendix}-\eqref{eq:source_terms} above, the main text proceeds to resolve the scattering coefficients for a given MG configuration by stipulating the boundary conditions discussed therein. As formulated in {Eq. \eqref{eq:matrix_system}, these boundary conditions can be resolved via matrix inversion, involving the unknown coefficient vector $\mathbf{I}$, the excitation vector $\mathbf{V}$, and the impedance matrix $\mathbf{Z}$. The entries of these vectors and matrix are formulated using the coefficients and overlap integrals derived above; for completeness, we present these explicitly here. Specifically, the unknown vector is composed of all field amplitudes is given by
\begin{equation}
\mathbf{I} = 
\begin{bmatrix}
\mathcal{R}_{\mathbf{n}}^{\text{TM}} \\
\mathcal{R}_{\mathbf{n}}^{\text{TE}} \\
\mathcal{T}_{\mathbf{n}}^{\text{TM}} \\
\mathcal{T}_{\mathbf{n}}^{\text{TE}} \\
\mathcal{A}_{+,\mathbf{m}}^{\text{TM},(i)} \\
\mathcal{A}_{+,\mathbf{m}}^{\text{TE},(i)} \\
\mathcal{A}_{-,\mathbf{m}}^{\text{TM},(i)} \\
\mathcal{A}_{-,\mathbf{m}}^{\text{TE},(i)}
\end{bmatrix},
\label{eq:I_vector}
\end{equation}
while the excitation vector $\mathbf{V}$ collects the source terms reads
\begin{equation}
\mathbf{V} = 
\begin{bmatrix}
-S_{\mathbf{n}}^{(\text{I})} \\
-S_{\mathbf{n}}^{(\text{II})} \\
0 \\
0 \\
-S_{\mathbf{m}}^{(\text{III},i)} \\
-S_{\mathbf{m}}^{(\text{IV},i)} \\
 0\\
0
\end{bmatrix},
\label{eq:V_vector}
\end{equation}
where the zeros correspond to the $z=-h$ interface equations Eqs. \eqref{eq:Ex_lower} and \eqref{eq:Ey_lower}, which carry no incident wave contribution. 

\begin{table*}[!t]
\caption{Modal Coefficients for Boundary Conditions}
\label{tab:coefficients}
\centering
\renewcommand{\arraystretch}{2.5}
\begin{tabular}{|c|c|c|c|c|c|}
\hline
& $\alpha^{\text{TM}(q)}_{\mathbf{n}}$& $\alpha^{\text{TE}(q)}_{\mathbf{n}}$& $\beta^{\text{TM}(q,i)}_{\mathbf{m}}$& $\beta^{\text{TE}(q,i)}_{\mathbf{m}^*}$& $S_{\mathbf{n}}^{(q)}$ or $S_{\mathbf{m}}^{(q,i)}$ \\
\hline
\hline
$q=\text{I}$& 
$-\dfrac{\eta k_x k_z}{k_x^2 + k_y^2}$& 
$-\dfrac{\eta k k_y}{k_x^2 + k_y^2}$ & 
$-\dfrac{j \eta_d^{(i)} k_{z,d}^{(i)} k_{x,d}^{(i)}}{[k_{x,d}^{(i)}]^2 + [k_{y,d}^{(i)}]^2}$&
$\dfrac{j \eta k k_{y,d}^{(i)} }{[k_{x,d}^{(i)}]^2 + [k_{y,d}^{(i)}]^2}$&
$s^{(\text{I})}\delta_{\mathbf{n},\mathbf{0}}$\\
\hline
$q=\text{II}$ & 
$-\dfrac{\eta k_y k_z}{k_x^2 + k_y^2}$& 
$\dfrac{\eta k k_x}{k_x^2 + k_y^2}$ & 
$-\dfrac{j \eta_d^{(i)} k_{z,d}^{(i)} k_{y,d}^{(i)} }{[k_{x,d}^{(i)}]^2 + [k_{y,d}^{(i)}]^2}$&
$-\dfrac{j \eta k k_{x,d}^{(i)}}{[k_{x,d}^{(i)}]^2 + [k_{y,d}^{(i)}]^2}$&
$s^{(\text{II})}\delta_{\mathbf{n},\mathbf{0}}$\\
\hline
$q=\text{III}$ & 
$\dfrac{\eta k k_y}{k_x^2 + k_y^2}$& 
$-\dfrac{\eta k_x k_z}{k_x^2 + k_y^2}$ & 
$\dfrac{j \eta_d^{(i)} \varepsilon_{d,r}^{(i)} k k_{y,d}^{(i)} }{[k_{x,d}^{(i)}]^2 + [k_{y,d}^{(i)}]^2}$ &
$\dfrac{j \eta k_{x,d}^{(i)} k_{z,d}^{(i)}}{[k_{x,d}^{(i)}]^2 + [k_{y,d}^{(i)}]^2}$&
$s^{(\text{III})}\chi^{*(i)}_{\mathbf{0},\mathbf{m}}$\\
\hline
$q=\text{IV}$ & 
$-\dfrac{\eta k k_x}{k_x^2 + k_y^2}$& 
$-\dfrac{\eta k_y k_z}{k_x^2 + k_y^2}$ & 
$-\dfrac{j \eta_d^{(i)} \varepsilon_{d,r}^{(i)} k k_{x,d}^{(i)} }{[k_{x,d}^{(i)}]^2 + [k_{y,d}^{(i)}]^2}$ &
$\dfrac{j \eta k_{y,d}^{(i)} k_{z,d}^{(i)} }{[k_{x,d}^{(i)}]^2 + [k_{y,d}^{(i)}]^2}$&
$s^{(\text{IV})}\psi^{*(i)}_{\mathbf{0},\mathbf{m}}$\\
\hline
\end{tabular}
\end{table*}

The rows of the impedance matrix $\mathbf{Z}$ are populated by the $\alpha$, $\beta$, $\psi$, and $\chi$ coefficients appearing in Eqs. \eqref{eq:Ex_upper}-\eqref{eq:Hy_lower}, organized according to the boundary condition type $q$ and mode indices $\mathbf{n}$, $\mathbf{m}$, and perforation index $i$.
Denoting $\tilde{\psi}^{(i)}_{\mathbf{n},\mathbf{m}} = \frac{\psi^{(i)}_{\mathbf{n},\mathbf{m}}}{L_xL_y}$, $\tilde{\chi}^{(i)}_{\mathbf{n},\mathbf{m}} = \frac{\chi^{(i)}_{\mathbf{n},\mathbf{m}}}{L_xL_y}$, and $\Gamma^{(i)}_{m_{x/y}} = \frac{d_x^{(i)}d_y^{(i)}}{4}(1+\delta_{m_{x/y},0})$, the matrix $\mathbf{Z}$ in Eq. \eqref{eq:matrix_system} is assembled in block form as Eq. \eqref{eq:Z_matrix}, where the block rows correspond to $q=\text{I,II}$ at $z=0$; $q=\text{I,II}$ at $z=-h$; $q=\text{III,IV}$ at $z=0$; and $q=\text{III,IV}$ at $z=-h$, and the block columns correspond to $[\mathcal{R}_{\mathbf{n}}^{\text{TM}}, \mathcal{R}_{\mathbf{n}}^{\text{TE}}]$, $[\mathcal{T}_{\mathbf{n}}^{\text{TM}}, \mathcal{T}_{\mathbf{n}}^{\text{TE}}]$, $[\mathcal{A}_{+,\mathbf{m}}^{\text{TM},(i)}, \mathcal{A}_{+,\mathbf{m}}^{\text{TE},(i)}]$, $[\mathcal{A}_{-,\mathbf{m}}^{\text{TM},(i)}, \mathcal{A}_{-,\mathbf{m}}^{\text{TE},(i)}]$. The blocks $\mathbf{A}$, $\mathbf{C}$, $\mathbf{F}^{\pm}$, $\mathbf{H}^{\pm}$ are diagonal, since the corresponding boundary condition equations Eqs. \eqref{eq:Ex_upper}-\eqref{eq:Hy_lower} contain no summation over mode indices, resulting in a one-to-one coupling between each equation and the same-index unknown. The blocks $\mathbf{B}^{\pm}$, $\mathbf{D}^{\pm}$, $\mathbf{E}$, $\mathbf{G}$ are full matrices, as the corresponding equations involve a summation over all mode indices, coupling each equation to all unknowns of the respective group.

\begin{equation}
\mathbf{Z} = 
\begin{bmatrix}
\mathbf{A} & \mathbf{0} & \mathbf{B}^+ & \mathbf{B}^- \\
\mathbf{0} & \mathbf{C} & \mathbf{D}^+ & \mathbf{D}^- \\
\mathbf{E} & \mathbf{0} & \mathbf{F}^+ & \mathbf{F}^- \\
\mathbf{0} & \mathbf{G} & \mathbf{H}^+ & \mathbf{H}^-
\end{bmatrix},
\label{eq:Z_matrix}
\end{equation}
with the block entries given by:
\begin{subequations}
\label{eq:Z_blocks}
\begin{align}
\mathbf{A} &= 
\begin{bmatrix} 
\alpha_{\mathbf{n}}^{\text{TM}(\text{I})} & \alpha_{\mathbf{n}}^{\text{TE}(\text{I})} \\
\alpha_{\mathbf{n}}^{\text{TM}(\text{II})} & \alpha_{\mathbf{n}}^{\text{TE}(\text{II})}
\end{bmatrix}, \label{eq:Z_A}\\
\mathbf{B}^{\pm} &= 
\begin{bmatrix} 
\mp\tilde{\psi}^{(i)}_{\mathbf{n},\mathbf{m}}\beta_{\mathbf{m}}^{\text{TM}(\text{I},i)} & 
-\tilde{\psi}^{(i)}_{\mathbf{n},\mathbf{m}}\beta_{\mathbf{m}}^{\text{TE}(\text{I},i)} \\
\mp\tilde{\chi}^{(i)}_{\mathbf{n},\mathbf{m}}\beta_{\mathbf{m}}^{\text{TM}(\text{II},i)} & 
-\tilde{\chi}^{(i)}_{\mathbf{n},\mathbf{m}}\beta_{\mathbf{m}}^{\text{TE}(\text{II},i)}
\end{bmatrix}, \label{eq:Z_B}\\
\mathbf{C} &= 
\begin{bmatrix} 
-\alpha_{\mathbf{n}}^{\text{TM}(\text{I})} & \alpha_{\mathbf{n}}^{\text{TE}(\text{I})} \\
-\alpha_{\mathbf{n}}^{\text{TM}(\text{II})} & \alpha_{\mathbf{n}}^{\text{TE}(\text{II})}
\end{bmatrix}e^{-jk_zh}, \label{eq:Z_C}\\
\mathbf{D}^{\pm} &= \mathbf{B}^{\pm}\, e^{\pm jk_{z,d}^{(i)}h}, \label{eq:Z_D}\\
\mathbf{E} &= 
\begin{bmatrix} 
\alpha_{\mathbf{n}}^{\text{TM}(\text{III})}\chi^{*(i)}_{\mathbf{n},\mathbf{m}} & 
\alpha_{\mathbf{n}}^{\text{TE}(\text{III})}\chi^{*(i)}_{\mathbf{n},\mathbf{m}} \\
\alpha_{\mathbf{n}}^{\text{TM}(\text{IV})}\psi^{*(i)}_{\mathbf{n},\mathbf{m}} & 
\alpha_{\mathbf{n}}^{\text{TE}(\text{IV})}\psi^{*(i)}_{\mathbf{n},\mathbf{m}}
\end{bmatrix}, \label{eq:Z_E}\\
\mathbf{F}^{\pm} &= 
\begin{bmatrix} 
-\Gamma^{(i)}_{m_y}\beta_{\mathbf{m}}^{\text{TM}(\text{III},i)} & 
\mp\Gamma^{(i)}_{m_y}\beta_{\mathbf{m}}^{\text{TE}(\text{III},i)} \\
-\Gamma^{(i)}_{m_x}\beta_{\mathbf{m}}^{\text{TM}(\text{IV},i)} & 
\mp\Gamma^{(i)}_{m_x}\beta_{\mathbf{m}}^{\text{TE}(\text{IV},i)}
\end{bmatrix}, \label{eq:Z_F}\\
\mathbf{G} &= 
\begin{bmatrix} 
\alpha_{\mathbf{n}}^{\text{TM}(\text{III})}\chi^{*(i)}_{\mathbf{n},\mathbf{m}} & 
-\alpha_{\mathbf{n}}^{\text{TE}(\text{III})}\chi^{*(i)}_{\mathbf{n},\mathbf{m}} \\
\alpha_{\mathbf{n}}^{\text{TM}(\text{IV})}\psi^{*(i)}_{\mathbf{n},\mathbf{m}} & 
-\alpha_{\mathbf{n}}^{\text{TE}(\text{IV})}\psi^{*(i)}_{\mathbf{n},\mathbf{m}}
\end{bmatrix}e^{-jk_zh}, \label{eq:Z_G}\\
\mathbf{H}^{\pm} &= \mathbf{F}^{\pm}\, e^{\pm jk_{z,d}^{(i)}h}. \label{eq:Z_H}
\end{align}
\end{subequations}
To ensure dimensional consistency of $\mathbf{Z}$, the boundary condition equations corresponding to the $H_x$ and $H_y$ tangential field components 
($q=\text{III}$ and $q=\text{IV}$) are multiplied by $\eta$ prior to assembly into Eq. \eqref{eq:matrix_system}, so that all elements of $\mathbf{Z}$ carry consistent impedance units.

\section{Fabrication-Related Adjustments}
\label{appendix:fabrication}
In this appendix, we list the perforation dimensions of the two fabricated dielectric-loaded MGs before and after the fabrication-related adjustments mentioned in Sections \ref{subsec:pbs} and \ref{subsec:dual}. The as-manufactured dimensions were obtained by introducing the corner rounding (with the radius of curvature specified by the mechanical workshop for this CNC process, approximately $0.5$~mm) into the full-wave model of the nominal design, which slightly shifted the efficiency peak away from the design frequency, followed by a brief optimization of the perforation dimensions to restore the peak at $20$~GHz. Since the shift was small, only small compensating changes were required; for completeness, we present in Tables \ref{tab:AppC1} and \ref{tab:AppC2}, respectively, the relevant geometrical dimensions of the prototypes of Sections \ref{subsec:pbs} and \ref{subsec:dual}, documenting these minor adjustments.

\begin{table}[H]
\renewcommand{\arraystretch}{1.15}
\caption{Perforation dimensions of the polarization beam splitter (Section \ref{subsec:pbs}) before (nominal, sharp corners) and after the fabrication-related adjustments (including corner rounding)}.
\label{tab:AppC1}
\centering
\begin{tabular}{lcc}
\hline
Parameter & Nominal design & After adjustment\\
\hline
$h$ [mm] & 6.3 & 6.39\\
$d_x^{(1)}$ [mm] & 4.64 & 4.75\\
$d_y^{(1)}$ [mm] & 7.0 & 6.98\\
$a_x^{(1)}$ [mm] & $-3.9$ & $-3.89$\\
$d_x^{(2)}$ [mm] & 6.9 & 6.78\\
$d_y^{(2)}$ [mm] & 7.3 & 7.4\\
$a_x^{(2)}$ [mm] & 2.8 & 2.87\\
$L_y$ [mm] & 9.9 & 10\\
\hline
\end{tabular}
\end{table}

\begin{table}[H]
\renewcommand{\arraystretch}{1.15}
\caption{Perforation dimensions of the dual-polarized anomalous refraction MG (Section \ref{subsec:dual}) before (nominal, sharp corners) and after the fabrication-related adjustments (including corner rounding)}.
\label{tab:AppC2}
\centering
\begin{tabular}{lcc}
\hline
Parameter & Nominal design & After adjustment\\
\hline
$h$ [mm] & 5.8 & 5.96\\
$d_x^{(1)}$ [mm] & 4.6 & 4.73\\
$d_y^{(1)}$ [mm] & 4.7 & 4.83\\
$a_x^{(1)}$ [mm] & $-3.98$ & $-3.9$\\
$d_x^{(2)}$ [mm] & 7.0 & 6.8\\
$d_y^{(2)}$ [mm] & 7.9 & 7.81\\
$a_x^{(2)}$ [mm] & 2.79 & 2.86\\
$L_y$ [mm] & 13.2 & 12.7\\
\hline
\end{tabular}
\end{table}

\bibliography{TransmissiveMetallodielectricMGs}

\end{document}